\documentclass{article}

\usepackage{arxiv}

\usepackage[utf8]{inputenc} 
\usepackage[T1]{fontenc}    
\usepackage{hyperref}       
\usepackage{url}            
\usepackage{booktabs}       
\usepackage{amsfonts}       
\usepackage{nicefrac}       
\usepackage{microtype}      
\usepackage{graphicx}
\usepackage{doi}
\usepackage{amsmath}

\title{Patient recruitment forecasting in clinical trials using
time-dependent Poisson-gamma model and homogeneity testing criteria
}

\date{}

\def\Gam{{\rm Ga}}

\def\Bin{{\rm Bin}}

\def\a{\alpha}
\def\be{\beta}

\def\e{\varepsilon}

\def\la{\lambda}
\def\La{\Lambda}

\def\de{\delta}

\def\d{{\rm d}}

\def\E{{\mathbf E}}
\def\Pr{{\mathbf P}}
\def\Va{{\mathbf {Var}}}

\def\cN{{\mathcal N}}

\def\what{\widehat}

\def\nn{\nonumber}

\newtheorem{lemma}{Lemma}[section]

\def\bee{\begin{equation}\label}
\def\ene{\end{equation}}
\def\beeq{\begin{eqnarray}\label}
\def\eneq{\end{eqnarray}}
\def\beeqn{\begin{eqnarray*}}
\def\eneqn{\end{eqnarray*}}
\def\beth{\begin{theorem}\label}
\def\enth{\end{theorem}}

\def\belem{\begin{lemma}\label}
\def\enlem{\end{lemma}}

\author{
\\
{\large Volodymyr Anisimov\thanks{E-mail: \texttt{vanisimo@amgen.com}}}
\\
	Data Science\\
	Center for Design \& Analysis\\
	Amgen, London, UK \\
	\And
	\\ {\large Lucas Oliver} \\
	Data Science\\
	Center for Design \& Analysis\\
	Amgen, London, UK \\
}

\renewcommand{\headeright}{}
\renewcommand{\undertitle}{}
\renewcommand{\shorttitle}{Recruitment forecasting using time-dependent Poisson-gamma model}

\hypersetup{
pdftitle={Enhanced patient recruitment forecasting in clinical trials using time-dependent
Poisson-gamma model and analysis of criteria for testing for homogeneity},
pdfauthor={Vladimir Anisimov, Lucas Oliver},
pdfkeywords={Forecasting patient recruitment, Poisson-gamma model, Time-dependence, Clinical trial},
}

\begin{document}
\maketitle

\begin{abstract}
Clinical trials in the modern era are characterized by their complexity and high costs
and usually involve
hundreds/thousands of patients to be recruited across multiple clinical centres in many countries,
as typically a rather large sample size is required in order to prove the efficiency of a particular drug.

As the imperative to recruit vast numbers of patients across multiple clinical centres has become a major challenge, an accurate forecasting of patient recruitment
is one of key factors for the operational success of clinical trials.
A classic Poisson-gamma (PG) recruitment model assumes time-homogeneous recruitment rates.
However, there can be potential time-trends in the recruitment driven by various factors, e.g. seasonal changes, exhaustion of patients on particular treatments in some centres, etc.
Recently a few authors considered some extensions of the PG model to time-dependent rates under some particular assumptions.
In this paper, a natural generalization of the original PG model to a PG model with non-homogeneous time-dependent rates is introduced.
It is also proposed a new analytic methodology for modelling/forecasting
patient recruitment using a Poisson-gamma approximation of recruitment processes in different countries and globally.
The properties of some tests on homogeneity of the rates (non-parametric one using a Poisson model and two parametric tests using Poisson and PG model) are investigated.
The techniques for modeling and simulation of the recruitment using time-dependent model are discussed.
For re-projection of the remaining recruitment
it is proposed to use a moving window and re-estimating parameters at every interim time.
The results are supported by simulation of some artificial data sets.
\end{abstract}

\section{Introduction}

Contemporary late-phase clinical trials involve
hundreds or even thousands of patients to be recruited across multiple clinical centres in many countries
as typically it is required to achieve a rather large sample size to prove the efficiency of a particular drug.
Citing Ken Getz, director, Tufts Center for Study of Drug Development, USA,
patient recruitment and retention are among the greatest challenges that
the clinical research enterprise faces today, and they are a major cause of drug development delays.

To address these challenges, novel approaches and predictive analytic techniques are required for efficient data analysis, monitoring, and decision-making.

Consider typical multicentre clinical trials used in pharmaceutical industry where the patients are recruited by many clinical centres in different countries and then randomized to different treatments.


Using Poisson models for modelling patient recruitment in clinical trials
is a well-accepted approach.
Several papers use Poisson processes in different clinical centres with fixed recruitment rates
(Carter et al., \cite{carter05}; 
Senn \cite{senn97,senn98}. 

However, in real trials different centres typically have different capacity and productivity,
and the recruitment rates vary.
To model this variation, Anisimov and Fedorov \cite{anfed07,anfed07_Nant}
introduced a Poisson-gamma (PG) model, where the patient arrival at clinical centres
is modeled using Poisson
processes with some rates and the variation in
rates among different centres is modeled using a gamma distribution.

This model can be described in the framework
of the empirical Bayesian approach where the prior distribution of the rates is a gamma
distribution with parameters that are evaluated either using historical data or data provided
by study investigators.
In \cite{anfed07} a maximum likelihood technique was proposed for estimating the parameters of the rates 
and the Bayesian technique for adjusting the posterior
distribution of the rates at any interim time using recruitment data in the individual centres.
Various applications to real trials
are considered in
\cite{anfed07,an-dow-fed07,an09a,an11a}.

This technique was developed further to account for random delays and closure of
clinical centres 
(Anisimov \cite{an-dow-fed07,an09a,an11a,an20})
and for modelling events in event-driven trials (oncology)
(Anisimov \cite{an11b,an20} and Anisimov et al. \cite{an-StRC22}).

Note that independently Gajewski et al. \cite{gajew-sim08} 
considered later a similar approach to modelling recruitment as a Poisson process
with gamma distributed rate, however their model is applicable to only one clinical centre.

The PG model was also used in Mijoule et al. \cite{savy12} 
to consider some
extensions related to other distributions of the recruitment rates and sensitivity analysis,
in Minois et al. \cite{savy17}
to evaluate the duration of recruitment process when historical trials are available,
and
in Bakhshi et al. \cite{baksenn13} 
for the evaluation of parameters of a PG model using
meta-analytic techniques for historic trials.
A survey on using mixed Poisson models is provided
in a discussion paper \cite{an16b}.

It is also worth mentioning the recent investigations
devoted to centralized statistical monitoring and
forecasting recruitment performance
and also forecasting patient recruitment
under various restrictions
and creating an optimal cost-efficient recruitment design, Anisimov and Austin
\cite{an-aus20,an-aus22a,an-aus23,an-23}.

There are also other approaches to recruitment modelling described in the literature, however,
they are mainly dealing with the analysis of the global recruitment
and therefore have some limitations.
These approaches typically require
rather large number of centres and patients 
and cannot
be applied to predicting recruitment on centre/country level
(see survey papers
by Barnard et al. \cite{barnard10}; Heitjan et al. \cite{heitjan15},
Gkioni et al. \cite{gkioni19}.

Currently, the originally developed PG model, \cite{anfed07,an11a},
gained world-wide recognition and in some recent papers is now called
"one of the most popular techniques" and an "industry-standard" model.

This model
assumes that patient recruitment rates do not change
over time. However, in real trials there can be some time-trends, seasonal changes,
and recruitment can be changing in some countries over time due to different reasons,
e.g. a slowdown in recruitment near the end of a trial, etc.

Therefore, to capture these situations, in some recent papers
\cite{Perevoz2022,lan-heitjan19,Perper2023,Savy2023,Armando2024,urban2022},
the standard PG model was extended
to the cases where the recruitment rates can be time-dependent.

Lan et al. \cite{lan-heitjan19} proposed a non-homogeneous Poisson process model that allows
for staggered centre activation and heterogeneity within centres is modeled by a gamma distribution
and assumes that after a period of steady recruitment,
the centre mean recruitment rate gradually declines as a negative exponential with some coefficient.
They calculate the posterior distribution and consider some applications to real trials.
Urbas et al. \cite{urban2022} proposed a more general time-dependent PG model that allows
for a wider range of recruitment rate functions.
The model is fitted using a maximum likelihood approach and is used to select the best model among a set of candidate models.
Perperoglou et al. \cite{Perper2023} compared the performance of the standard PG model to two time-dependent models:
the models proposed in \cite{lan-heitjan19} and \cite{urban2022}.
The study found that the time-dependent models outperformed the PG model in terms of prediction accuracy,
especially in trials with time-varying recruitment rates.
Best et al. \cite{Perevoz2022} considered a standard PG model as the starting point and
developed a more flexible version that introduces variation in rates over time as a function of COVID-19-dependent
covariates which is implemented in a fully Bayesian probabilistic framework.
Turchetta et al. \cite{Savy2023} proposed a time-dependent PG model
allowing recruitment rates to vary over time through B-splines that is fitted using a Bayesian approach.
The model is evaluated in a simulation study and found to perform well in a variety of scenarios.
In another paper, Turchetta et al. \cite{Armando2024} consider predictive performance of the proposed technique in forecasting the recruitment process of two HIV vaccine trials.

All six papers above contribute to the advancement of recruitment modelling in multicentre clinical trials
by extending the standard time-homogeneous PG model to accommodate time-dependent recruitment rates.

The flexibility of the original PG model allows for rather natural extension to model time-dependent rates
on a centre level. However, the main question here is: what type of time-dependence to use?
Also, how to detect the time-dependent behavior of the rates, and how to test for
this dependence in real trials.
Thus, the choice of a suitable time-dependent model is trial specific.

In this paper, the authors consider a natural generalization of a PG model
to a time-dependent model using a general framework of a standard PG model \cite{anfed07,an11a}
which is similar to the way considered in \cite{urban2022}.
The basic methodology of using a PG model with time-dependent rates
is developed by the first author in \cite{an-24}. This paper
extends some results of \cite{an-24}, in particular,
we investigate two new parametric criteria for testing the recruitment rates for homogeneity,
a Poisson criterion and a PG one in addition to a non-parametric Poisson
criterion \cite{an-24}.
In \cite{an-24} it is also proposed a novel analytic methodology for modelling/forecasting
patient recruitment using the recent results in \cite{an-aus20}
on a PG approximation of the sums of PG processes in different centres
that can be applied to modelling recruitment even in not so large countries/regions,
which is also described in this paper for completeness of presentation.

Note that this analytic methodology
is not reflected in the six papers mentioned above
but is crucial as the country process under general assumptions is not a PG process
and also for a small number of centres
a normal approximation for the total recruitment process in general cannot be applied.
Therefore, for using analytic tools for modelling/forecasting
patient recruitment on different levels instead of Monte Carlo simulation,
we need to develop some analytic approximations of country
processes for time-dependent models similar to \cite{an-aus20}, as well.

The first author also proposed two new criteria,
a non-parametric criterion and the parametric one
using Poisson assumptions for testing
the recruitment rates for time-dependence, Section \ref{Pois-test},
and the fast algorithm of simulation
of a time-dependent model in R using discrete event simulation, Section \ref{Simulation},
that is used in implementations.

The new parametric criterion for testing
rates for time-dependence using assumptions of a PG model, Section \ref{PG-test},
is analyzed by the second author who also ran many simulated scenarios using Monte Carlo technique to test the applicability of non-homogeneous PG model for prediction and interim re-projection
of recruitment and compare
with using a classic PG model. Some examples are provided in Section \ref{Sim-examples}.
These results are briefly discussed in Section \ref{Implem}.

\section{Poisson-gamma model with homogeneous rates}\label{sec2}

Consider a multicentre clinical trial which is designed
to recruit $n$ patients by $N$ clinical centres.
A standard PG model, \cite{anfed07,an11a},
assumes that the recruitment rates in different centres are modeled
as independent gamma distributed random variables
with some parameters that do not depend on time.
Thus, the recruitment process in any interval
where the centre is active follows a time-homogeneous
mixed Poisson process with gamma distributed rate (PG process).

First consider some definitions.
Denote by $\Pi(a)$ a Poisson random variable with parameter $a$, and by
$\Pi_a(t)$ an ordinary homogeneous Poisson process with rate $a$.
This means, for any $t>0$,
$$
\Pr(\Pi_a(t) = k) = e^{-a t}\frac{(at)^k}{k!}, \ k=0,1,\dots
$$
Also denote by ${\rm Ga}(\alpha,\beta)$ a gamma
distributed variable with parameters $(\alpha,\beta)$ (shape and rate) and
probability density function
\begin{equation}\label{e00}
f(x,\alpha,\beta) =
\frac {e^{- \beta x} \beta^{\alpha} x ^{\alpha-1} }{ \Gamma(\alpha)},\
x >0,
\end{equation}
where
$ \Gamma(\alpha) = \int_0^\infty e^{- x}  x ^{\alpha-1} {\rm d}x$ is a gamma function.

\subsection{The case of homogeneous rates}

Consider first the basic definitions used in a standard PG model as some similar definitions
will be used in PG model with time-dependent rates.
Let $\Pi_\lambda(t)$ be a doubly stochastic Poisson process
where $\lambda={\rm Ga}(\alpha,\beta)$.
According to terminology in \cite{bernardo04},
this is a PG process with parameters $(t,\alpha,\beta)$.
Denote it as $\text{PG}(t,\alpha,\beta)$ to reflect the dependence on $(\a,\be)$.
Then for any $k = 0,1,\dots$,
\begin{equation}\label{PG}
\Pr(\text{PG}(t,\alpha,\beta) =  k)
= \frac{\Gamma(\alpha + k)}{k!\ \Gamma(\alpha)}\ \frac{t^{k}\beta^{\alpha}}
{(\beta + t)^{\alpha + k}}.
\end{equation}

For $t = 1$, $\Pi_{\lambda}(1)$ has the same distribution as $\Pi(\lambda)$,
which is a PG variable with parameters $(\alpha,\beta)$.
For this case we omit $t=1$ and use the notation $\text{PG}(\alpha,\beta)$ instead of $\text{PG}(1,\alpha,\beta)$.

Note that
$\Pi_{\lambda}(t)$
has a
negative binomial distribution and for any $\alpha >0, \beta > 0, t >0 $, the following relation is true:
for any $k = 0,1,\dots$,
\begin{equation}\label{NB1}
\Pr(\text{PG}(t,\alpha,\beta) =  k) = \Pr(\text{Nb}( \alpha,\beta/(\beta + t)) = k) 
\end{equation}
where $\text{Nb}( \alpha,p)$ denotes a random variable
which has a negative binomial distribution with size $\alpha$  and
probability $p$:
$$
\Pr(\text{Nb}( \alpha,p)= k) = \frac{\Gamma(\alpha + k)}{k!\ \Gamma(\alpha)}p^{\alpha}{(1 - p)}^{k},\ k = 0,1,\dots
$$
This relation is useful in calculations of a PG distribution using R, as there are
standard functions pnbinom(), dnbinom().

Now consider modelling recruitment.
Denote by $n_i(t)$ the recruitment process in centre $i$ (the number of patients recruited
in time interval $[0,t]$), and by $n(I_j,t)$ the recruitment process in some region
with the set $I_j$ of $N_j$ clinical centres.
Also denote by $u_i$ 
the date of initiation of centre $i$. 

Assume that the recruitment process in centre $i$ during an active recruitment stage
 follows a homogeneous PG process
with rate $\lambda_i$
which has a gamma distribution with parameters $(\a_i,\be_i)$.
Then the mean rate $m_i$ and the variance $s_i^2$ are calculated as
$m_i = \a_i/\be_i$, $s_i^2 = \a_i/\be_i^2$.
Correspondingly, the instantaneous rate at time $t$ is
$\la_i(t) = \la_i \chi( u_i < t )$, where $\chi(A)$ is the indicator function of the event $A$.

Consider a more convenient representation via a cumulative rate.
Denote $x(t,u) = \max(0,t-u)$ (the duration of active recruitment at time $t$
for a centre activated at time $u$).
So, if centre $i$ is active at time $t > u_i$,
then $x(t,u_i) = t-u_i$.

Then the cumulative rate of the process $n_i(t)$
is $\La_i(t,u_i) = \la_i x(t,u_i)$.
This means,
if $\la_i=\Gam(\alpha_i,\beta_i)$,
$n_i(t)$ is a PG process with parameters $(x(t,u_i), \a_i, \be_i)$,
and the distribution of $n_i(t)$ can be calculated using (\ref{PG})
where in the right-hand side we should use $x(t,u_i)$ instead of $t$,
and parameters $(\alpha_i,\beta_i)$.

Correspondingly, for any region with a subset of centres $I_j$, the recruitment process $n(I_j,t)$ in this region
as a sum of the recruitment processes in the centres from this region
is a nonhomogeneous Poisson process
with the instantaneous rate $\lambda(I_j,t) = \sum_{i \in I_j} \lambda_i(t)$
and, in distribution,  $n(I_j,t) = \Pi(\Lambda(I_j,t))$, where the region
cumulative rate has the form
$$
\Lambda(I_j,t) = \sum_{i \in I_j}\Lambda_i(t,u_i).
$$
Note that in general a sum of gamma distributed variables with different
parameters doesn't have a gamma distribution, therefore,
in general, the process $n(I_j,t)$ is not a PG process.
The technique for the approximation of $n(I_j,t)$
by a PG process was developed in \cite{an-aus20}.

\section{Non-homogeneous PG model}\label{Non-Hom}

Consider now an extension of the classic PG model to time-dependent rates. Assume that the recruitment rates can proportionally change over time
according to some non-negative function $r(t), t \ge 0$.

A realistic type of dependence can be a piecewise linear dependence:
linear increase in some initial interval, then a constant rate in some larger interval
in the middle of recruitment,
and then linear decrease in the last interval closer to the end of recruitment.
Another opportunity for function $r(t)$ can be an exponential decay starting from 1 until some proportion, say, 50\%,
at the end of recruitment.

Of course the main question can arise - how to define the form of function $r(t)$.
However, this question is for separate investigation, and potentially the form of function $r(t)$ can be chosen
using similar historic trials, or monitoring the data for the current study
and use some extrapolations of the rate dependence up to the interim point
for the remaining interval.

Assume also for simplicity that the function $r(t)$ is the same for all centres,
that means, time-dependence is defined only by the duration of the recruitment in the trial.
However, in general, this assumption can be relaxed and we can assume different dependence
in different regions, though regional type dependence would be harder to extract from historic trials. Therefore, a non-homogeneous PG model can be defined according to \cite{an-24} as follows.

Assume that if a centre $i$ is active at time $t$, then the recruitment rate has the form
\bee{rate1}
\la_i r(t)
\ene
where $\la_i$ is some baseline rate which has a gamma distribution
with parameters $(\alpha_i,\beta_i)$, and $r(t)$ is some non-negative function. Then, if a centre $i$ is initiated at time $u_i$,
the instantaneous rate at time $t$ is
$\la_i(t) = \la_i r(t) \chi( u_i < t )$.

Define the cumulative proportional change in the rate for centre $i$ in the interval $[a,b]$ as
$$
R_i(a,b,u_i) =  \int_a^b r(x) \chi( u_i < x ) \d x
$$

In particular, if no change in time, $r(x) \equiv 1$, and  $R_i(a,b,u_i)$
is just a duration of recruitment window in the interval $[a,b]$.

The cumulative rate for centre $i$ in the interval $[a,b]$ is
$
\La_i(a,b,u_i) =  \la_i R_i(a,b,u_i).
$
Correspondingly, the recruitment process $n_i(t)$ in interval $[0,t]$ is again
a doubly stochastic PG process
with cumulative rate $\La_i(0,t,u_i)$.
Then,
$$ 
\E [\La_i(0,t,u_i)] = m_i  R_i(0,t,u_i), \\
\ \Va [\La_i(0,t,u_i)] = s_i^2  R_i^2(0,t,u_i)
$$ 

Consider now the recruitment process in some country $s$ with the set of centres $I_s$.
Then a country process
$$
n(I_s,t) = \sum_{i\in I_s} n_i(t)
$$
has a cumulative rate
$$
\La(I_s,t) = \sum_{i\in I_s} \La_i(0,t,u_i)
$$

Note that $n(I_s,t)$ is a mixed Poisson process but in general not a PG process.

Denote the mean and variance of $\La(I_s,t) $ as
\beeq{e20}
\begin{aligned}
E(I_s,t ) &=& \sum_{i\in I_s} m_i  R_i(0,t,u_i),  \\
S^{2}(I_s,t) &=&  \sum_{i\in I_s} s_i^2  R_i^2(0,t,u_i)
\end{aligned}
\eneq

Assuming that $S^{2}(I_s,t) >0$, let us introduce the variables
\bee{e21}
A(I_s,t) = E^{2}(I_s,t) / S^{2}(I_s,t),
B(I_s,t) = E(I_s,t) / S^{2}(I_s,t)
\ene

The following statement is an extension of the result in \cite{an-aus20}
to the case where the rates are gamma distributed with different parameters.

\belem{Lem1}
The distribution of $n(I_s,t)$
can be well approximated by
the distribution of a PG random variable
$PG(A(I_s,t),B(I_s,t))$.
\enlem

In \cite{an-aus20} it is shown using numerical calculations that
this approximation provides a very good fit even for a small number of
centres, $N =2,3$, and with a larger number of
centres the difference between
the exact and approximative distributions is decreasing
to a negligible value around $10^{-4}$ or less.

This result can be used to create the approximate values of
the mean and predictive bounds for the recruitment process
$n(I_s,t)$ at any time point $t$.

Indeed, by definition,
\bee{mean}
\E [n(I_s,t)] = E(I_s,t)
\ene
and $P$-confidence predictive bounds can be calculated using a quantile function
of the negative binomial distribution
\begin{verbatim}
qnbinom(P,size=M^2/S2,prob=M/(M+S2))
\end{verbatim}
where
at time $t$ we set
$M =E(I_s,t) $, $S2 = S^{2}(I_s,t)$.

To create predictions for the global recruitment we just need to
use in relation (\ref{e20}) the sum by all active centres.

This technique allows for creation of analytic predictions of the recruitment process
$n(I_s,t)$ over time with mean and predictive bounds assuming that the type
of time dependence, function $r(t)$, is known.

\subsection{Estimation at the interim stage}\label{Time-dep-estim}

The technique for estimating parameters using interim data
may depend on the form of time-dependence.

Using an assumption of a standard PG model \cite{anfed07,an11a},
where it is assumed that the parameters $(\a_i,\be_i)$ of the rates
are the same for all centres,
and also assuming that the function $r(t)$ is known,
for estimating parameters $(\a,\be)$ we can use the same
maximum likelihood technique developed in \cite{anfed07,an11a} where in calculations of the likelihood
we need to use the values $R_i(0,t,u_i)$ instead of recruitment windows $v_i$.
The posterior distribution of the rates can be calculated similarly as in \cite{anfed07,an11a}.

If the function $r(t)$ depends on some additional parameters,
then the technique of estimation will depend on the form
of the function $r(t)$.
In this case we can write a likelihood expression similarly as in \cite{anfed07,an11a}
and use $2+d$ optimisation, where $d$ is the number of parameters used
in the definition of the function $r(t)$.

Note that for time-dependent models, to closely capture the dependence on time,
it can be recommended to use a moving window including data only for the last several months.
The choice of the length of the window is trial specific,
but as a general recommendation, the number of centres and the number
of patients recruited during this window should be large enough
to apply a maximum likelihood technique in the framework of a PG model.
In applications to late-stage clinical trials, the length of the window can be chosen as 2,3,4 months.

\subsection{Simulation of non-homogeneous PG-model}\label{Simulation}

In general, it is preferable to use an analytic technique described
in Section \ref{Non-Hom}, Lemma \ref{Lem1}.
However,
there can be different restrictions
on the recruitment, or it can be required to evaluate different metrics related to recruitment performance. For example, to evaluate the probability that the recruitment process will fall into some region, say, within 20\% bounds from the predicted mean during a particular number of months, and
it's hard to develop the analytic relations for these situations.
Therefore, in these cases, Monte Carlo simulation can be the most appropriate way.
In addition, simulation can serve also as a useful tool to verify different analytic approaches.

Here we describe the technique of simulation which is based on a discrete event simulation approach and used directly in applications.


\subsubsection{Algorithm of simulation using R}\label{Simul}

Consider a trial with $N$ centres.
Assume that for a centre $i$ the time of initiation is $u_i$.
Denote $vecu = (u_1,\dots,u_N)$.

Consider also a given interval of simulation, $[0,T]$, which should be large
enough to ensure that for nearly all simulated trajectories the stopping point will be reached.
$T$ can be evaluated as the upper predictive bound of the recruitment time using some analytic expressions,
similar to \cite{an11a}.

Define vectors of parameters: \\
$veca = (\a_1,\dots,\a_N)$, $vecb = (\be_1,\dots,\be_N)$.
\vspace{-0.2cm}

\begin{enumerate}

\item Create a sequence $vecr = r(1:T)$ using function $r(t)$

\item Generate a vector of gamma distributed rates in centres: \\
$vecla$ = rgamma($N$, shape = $veca$, rate = $vecb$)

\item Define an output matrix for one run: \\
$outmat$ = matrix(nrow = $N$, ncol = $T$)

\item Use a loop by the number of centres $N$:
\end{enumerate}

\begin{itemize}
\item[--] For a centre $i$, create the vector \\
$veclaik$ = $vecla[i]$*$vecr$  \\
of the rate in centre $i$ for each day

\item[--] Update this vector by setting  \\
$veclaik[k] = 0$ for $k < vecu[i]$

\item[--] Simulate the numbers of recruited patients in each day $k = 1,\dots,T$
using vector of rates and Poisson distribution, \\
$vecptsday$ = rpois($T$, $veclaik$)

\item[--] Use cumsum($vecptsday$) to generate a cumulative number of patients in centre $i$
over time

\item[--] Save $outmat[i,]$ = cumsum($vecptsday$)
\end{itemize}

Now as the output in one simulation run, we get a matrix $outmat$ of the simulated
trajectories of the recruitment in each centre.

Depending on the purpose, we can use matrix calculations
to calculate country or global trajectories and calculate different characteristics,
e.g. time to reach a total sample size, time to reach country targets, etc.

By repeating this step many times, say, $10^4$ simulation runs,
we can calculate different summary statistics.

A computing time is quite small. To calculate
the trajectories of the mean, median and predictive bounds for the global recruitment
over time using $10^4$ runs, it takes seconds for $N \le 100$.
For these cases, the predictions using analytic expressions or
simulation practically coincide.

\section{Testing the recruitment rates for homogeneity}\label{Testing}

Consider a trial with $N$ centres initiated at different times $\{ u_i, i = 1,\dots,N \}$ and
during the recruitment period consider two disjoint intervals
$[a,b]$ and $[c,d]$ (in months) where $a < b < c < d$.
Define the observed number of patients in each interval
$n[a,b]$ and $n[c,d]$ and denote
the total number of patients
$n[a,b,c,d] = n[a,b]+n[c,d]$.

For a given centre, define the duration of the active recruitment window in the interval $[y,z]$
if the centre is initiated at time $u$ as the following function:
\begin{equation}\label{e2}
u(y,z,u) =  \left \{ \begin{array}{lll}
z - y & \hbox{if} & u < y \\
u - y & \hbox{if} & y \le u < z \\
0 & \hbox{if} &  u \ge z
        \end{array}
        \right.
\end{equation}

Denote by $U[a,b]$ and $U[c,d]$ the total durations
of recruitment windows in each interval,
that means,
\bee{Uabcd}
U[a,b] = \sum_i u(a,b,u_i); \ U[c,d] = \sum_i u(c,d,u_i)
\ene
and put
$U[a,b,c,d] = U[a,b]+U[c,d]$.

\subsection{Poisson-type test}\label{Pois-test}

Let us propose a non-parametric test using Poisson assumptions. Assume here a Poisson recruitment model.

Consider testing the hypothesis:

{\boldmath $H_0$:} The mean recruitment rates
in both intervals are the same.

\belem{lem2}
Given the total number of patients $n([a,b,c,d])$ and hypothesis {\boldmath $H_0$},
the number of patients $  n([a,b])$
has a binomial distribution $Bin(n([a,b,c,d]), p(a,b))$ where
$$
p(a,b) = \frac{U[a,b] }{U[a,b,c,d]}
$$
\enlem

{\em Proof.}
Under our assumptions, the numbers of patients
$n([a,b])$ and $  n([c,d])$
recruited in intervals $[a,b]$ and $ [c,d]$, respectively, are independent
Poisson random variables with rates $m U[a,b] $ and $m U[c,d]$.
Therefore, the proof follows from the known fact for Poisson variables.
Indeed, consider two independent Poisson variables
$\Pi(a_1)$ and $\Pi(a_2)$ with parameters $a_1$ and $a_2$.
Then the conditional distribution of  $\Pi(a_1)$, given 
$\Pi(a_1)+\Pi(a_2) = n$, is a binomial distribution
$\Bin(n,p)$ with
$p = a_1/(a_1+a_2)$.

\subsection{ Criterion for testing hypothesis {\boldmath $H_0$}}

Define two $P$-values: \\
Upper $P$-value
\bee{UppP}
P_{Upp}(a,b,c,d) = \Pr ( \Bin(n([a,b,c,d]), p(a,b)) \ge  n([a,b]))
\ene
Lower $P$-value
\bee{LowP}
P_{Low}(a,b,c,d) = \Pr ( \Bin(n([a,b,c,d]), p(a,b)) \le  n([a,b]))
\ene

Consider some critical level, say, $\de = 0.1 $.
Then: \\
1) If $P_{Upp}(a,b,c,d) \le \de$, with high probability the number of
patients $n([a,b])$ is unusually large and it is likely that
the mean recruitment rate in interval $[a,b]$ is larger compared to interval $[c,d]$.
\\
2) Otherwise, the mean rate in interval $[a,b]$ is smaller.


Correspondingly, we can detect intervals where the hypothesis {\boldmath $H_0$} is not true.
This allows us to detect intervals with rather high or rather low recruitment rates.

Note that this test is non-parametric as it does not include the rate $m$ which is beneficial.

The proposed approach is exact, it does not require estimation of the parameters,
and can be used for any number of centres and any number of recruited patients in any intervals.

As the next stage, we are coming to the following question.
How many centres do we need to detect a particular proportional difference in rates with a given confidence?

\subsubsection{Dependence of $P$-values on the number of centres and duration of the intervals.}\label{Depend}

Let us analyze the dependence of the $P$-values defined in (\ref{UppP}),(\ref{LowP}) on the
durations of active recruitment windows and the number of centres
for a time-dependent model.

Suppose that the rates for each centre in each interval $[a,b]$ and $[c,d]$
are the same and are equal to $m_1$ in interval $[a,b]$ and $m_2$ in interval $[c,d]$, respectively.
Let us analyze the dependence of $P_{Upp}$ on the values of rates
and $U[a,b]$ and $U[c,d]$.

This will allow to evaluate the number of centres needed for testing a particular difference in the rates
$m_1, m_2$.

For simplicity of notation, let us use index $j$ to define a particular interval
($j = 1$ corresponds to $[a,b]$, $j=2$ corresponds to $[c,d]$).
Correspondingly, denote by $U_j$ the total duration of recruitment windows
in interval $j$ defined in (\ref{Uabcd}).

Then the number of recruited patients $n_j$ in interval
$j$ is a Poisson random variable,
  $\pi_j = \Pi(m_jU_j)$,
and the expression for upper $P$-value can be written as
\bee{UppP-2}
P_{Upp} = \Pr ( X \ge 0)
\ene
where
$X = \Bin(\pi_1+\pi_2, p) - \pi_1$,
 $p=U_1/(U_1+U_2)$.

Then the mean of $X$ is calculated as:
\begin{equation}\label{Mean}
\begin{aligned}
\E[X] &=& \E[\Bin(\pi_1+\pi_2,p)] - \E[\pi_1] = \E[\pi_1+\pi_2]p - \E[\pi_1] \\
&=& 
m_1 U_1 (p-1) + m_2U_2 p
= \frac{U_1U_2}{U_1+U_2}(m_2-m_1)
\end{aligned}
\end{equation}

To calculate the variance, note that  $\pi_1$ and $\pi_2$ are independent.
Then, using the formula of total variance, we get
\beeq{TotVar}
\begin{aligned}
& \Va[X] = \E[\Va [ X \mid n_1,n_2 ]] + \Va[\E[X \mid n_1,n_2]] \\
& = \E[(n_1+n_2)p(1-p)] + \Va[n_1(p-1)+n_2 p] \\
& = (m_1U_1+m_2U_2)p(1-p) + m_1U_1(p-1)^2 + m_2U_2p^2 \\
& = \frac{U_1U_2}{U_1+U_2}(m_1+m_2)
\end{aligned}
\eneq

Note that in the real trials, the expressions for the mean and the variance of $X$
are typically rather large as $U_1$ and $U_2$ are large.
Therefore,
we can consider a normal approximation for $X$ in the form
\bee{NormApp}
X \approx \E [X] + \sqrt{\Va[X]} Z
\ene
where $Z = \cN(0,1)$ is a standard normal random variable.
Then
\beeq{Pupp2}
\begin{aligned}
 P_{Upp} & = \Pr(X \ge 0) \approx \Pr\left( Z \ge \frac{\frac{U_1U_2}{U_1+U_2}(m_1-m_2)}{\sqrt{\frac{U_1U_2}{U_1+U_2}(m_1+m_2)}} \right) \\
& = \Pr \left( Z \le \sqrt{\frac{U_1U_2}{U_1+U_2}}\frac{m_2-m_1}{\sqrt{m_1+m_2}}\right)
\end{aligned}
\eneq

Note that if $m_1 > m_2$, then $P_{Upp} \to 0$ as $U_1,U_2 \to \infty$.
Likewise, if $m_1<m_2$, $P_{Upp} \to 1$ as $U_1,U_2 \to \infty$.

We investigate the properties of this test when $m_1 > m_2$ (the rate is decreasing).

Note that the relation
$
P_{Upp} = \de
$
can be approximated by the relation
\bee{Quant0}
\sqrt{\frac{U_1U_2}{U_1+U_2}}\frac{m_2-m_1}{\sqrt{m_1+m_2}} = z_{\delta}
\ene
where $z_{\delta}$ is a $\de$-quantile of a standard normal distribution.

This relation has a general form and includes the total durations of recruitment windows $U_1, U_2$.
To analyze the dependence on the number of centres and durations of intervals, consider
a specific case when all centres are active in each interval, and there are $N_j$ active centres in interval
$j$.

In this case $U_j = L_j N_j, j=1,2$. Correspondingly,
\bee{EX}
\E[X] = \frac{N_1 L_1 N_2 L_2}{N_1 L_1 + N_2 L_2}(m_2-m_1),
\Va[X] = \frac{N_1 L_1 N_2 L_2}{N_1 L_1 + N_2 L_2}(m_1+m_2)
\ene
and the relation (\ref{Quant0}) has the form
\bee{Quant}
\sqrt{\frac{N_1L_1N_2L_2}{N_1L_1+N_2L_2}}\frac{m_2-m_1}{\sqrt{m_1+m_2}} = z_{\delta}
\ene

In particular, if $N_1=N_2=N$,  $L_1=L_2=L$
and $m_2 = qm_1$ for $q \in (0,1)$
(the rate has decreased from interval $1$ to interval $2$),
then relation  (\ref{Quant}) implies
\bee{NN}
N = \frac{2 z_{\delta}^2}{m_1 L}\cdot\frac{1+q}{(1-q)^2}
\ene

Thus, 
$N = O(1/(1-q)^2)$,
and detecting small differences in rates requires many clinical centres,
or we need to increase the duration of the interval $L$.

\begin{figure}
\centering
\includegraphics[width=12.0cm,height=7cm]{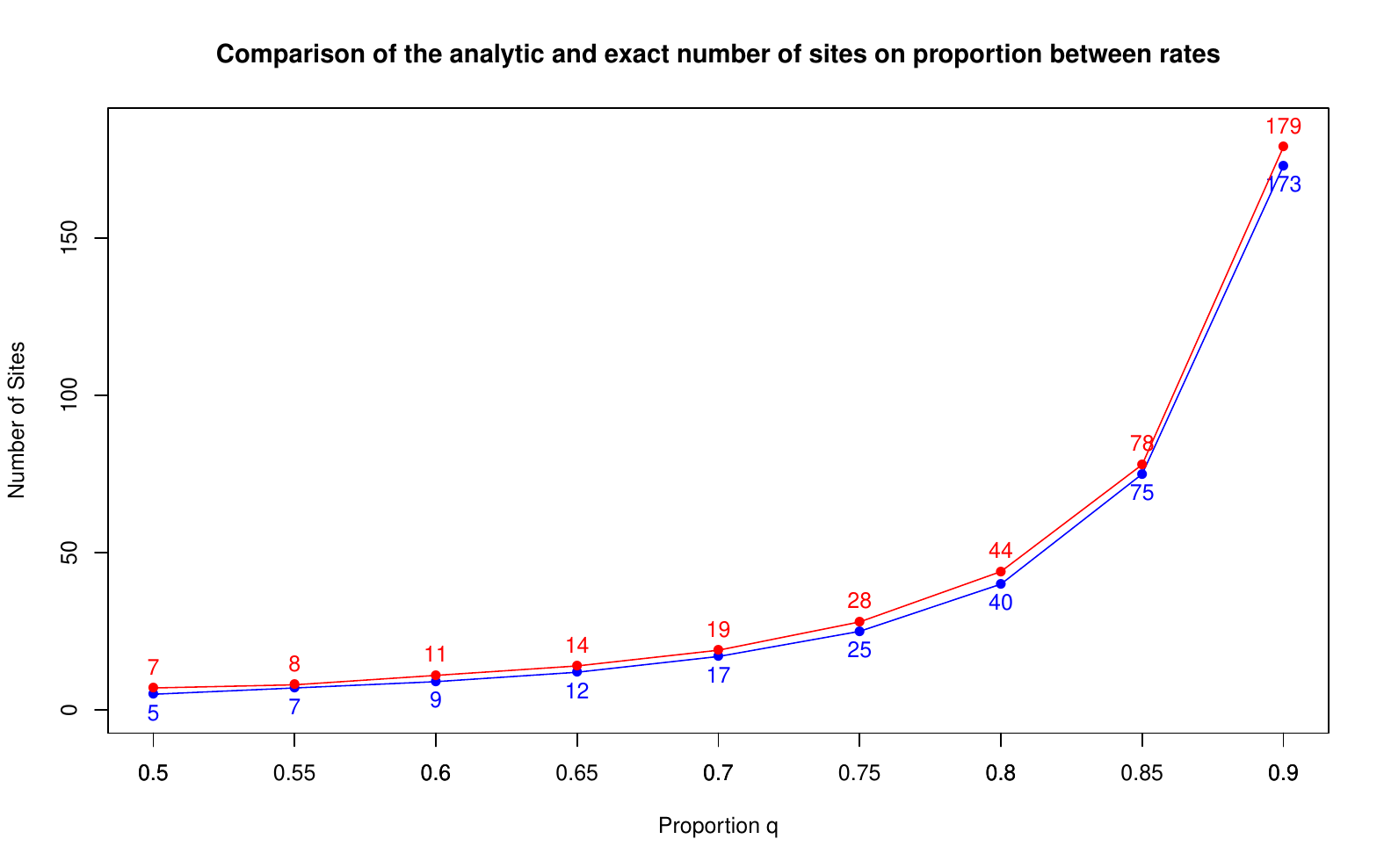}
\caption{Dependence of the number of centres/sites $N$ on the proportion $q$
in the rates. Analytic results are shown by a blue line and simulation results
by a red one.}
\label{fig1}
\end{figure}

Consider a numerical example.
Let's put $\delta=0.1$, $m_1=0.04$ (patients per centre per day), $L=90$ (days) and $q \in (0.5,0.9]$.
Figure \ref{fig1} shows how many centres are required to detect the proportion
in the rates $q$ with confidence level 90\% ($P_{Upp} \le 0.1$).
If $q$ is closer to 1, the number of centres is growing
as $1/(1-q)^2$.

In Figure \ref{fig1}, analytic results based on the normal approximation
are shown by a blue line.
The curve for the number of centres obtained by
using Monte Carlo simulation which is performed according to Section \ref{Simulation1}
is shown by a red line. Note that both curves nearly coincide.
That means, for evaluating the performance of Poisson test 
we can use analytic results based on simpler relation (\ref{NN}).

\subsection{The power of a Poisson test}\label{PoisPower}

The power of a Poisson test can be evaluated using Monte Carlo simulation.
We consider here a special case where $N_1=N_2=N$,  $L_1=L_2=L$
and $m_2 = qm_1$ for $q \in (0,1)$.

$P$-values are defined as in (\ref{UppP}), (\ref{LowP}) and for our case have the form
\beeq{PowerEq1}
\begin{aligned}
P_{Upp} &=& \Pr(\Bin(n,p) \ge n_1)  \\
P_{Low} &=& \Pr(\Bin(n,p) \le n_1)
\end{aligned}
\eneq
where $n=n_1+n_2$, $n_j$ is the number of recruited patients in interval $j$, and
$p= N_1L_1/(N_1L_1+N_2L_2)$.
\\

{\bf Criterion using $P$-values:} \\
For a given confidence level $\de$,\\
if $P_{Upp} \le \de$, we assume that $m_1 > m_2$; \\
if $P_{Low} \le \de$, we assume that $m_1 < m_2$. \\

Let us investigate the dependence of these quantities as random variables
depending on the observed data $(n_1,n_2)$.

Consider the approach for testing both hypotheses and analyzing the properties of this test and its power.


Consider first the upper $P$-value and given some hypothesis $H$ define a statistic
\bee{PowerEq2}
T_{H} = T_H(n_1,n_2) = \Pr_H(\Bin(n_1+n_2,p) \ge n_1 \mid (n_1,n_2) )
\ene
where $\Pr_{H}()$ is $P$-value but now it is considered as a random function depending on $(n_1,n_2)$.

Consider testing two hypotheses\\
1. $H_0$: $m_1=m_2=m$ \\
2. $H_1$: $m_2=qm_1$ where $q \in (0,1)$

Define the probability of a Type I error $\de$ (usually it is denoted
as $\a$, but $\a$ is already reserved for the shape parameter of a gamma distribution).

Then, using the hypothesis $H_0$, we need to define a significance level $a(\de)$ (critical region)
such that for a given $\de$,
\bee{PowerEq3}
\Pr (T_{H_0}(n_1,n_2) \le a(\de) ) = \de
\ene

It is known that under hypothesis $H_0$, for a continuous distribution,
$P$-value has a uniform distribution,
and this is also confirmed by simulations for our model.
Therefore, as the initial approximation we can choose $a(\de) = \de$.
However, in our case we have a discrete distribution,
thus, a more precise evaluation of the value $a(\de)$ can be done using
Monte Carlo simulation with up to $10^6$ runs.

Consider now the hypothesis $H_1$ and define the power of the test as
\bee{PowerEq4}
\Pr (T_{H_1}(n_1,n_2) \le a(\de) ) = 1-\e
\ene
where $\e$ is a probability of a Type II error, known as a "false negative".
The value $1-\e$ is a probability of correctly rejecting the null hypothesis $H_0$ given $H_1$.

Consider now the behavior of the power for some cases of
$m_i$, $N_i$ and $L_i$ using Monte Carlo simulation. Note that $ \E [T_{H_1}(n_1,n_2) ] = P_{Upp}$
as defined in (\ref{PowerEq1}).
Therefore, when we use the criterion
\bee{PowerEq5}
P_{Upp} \le a(\de)
\ene
to detect a difference in rates, if the relation (\ref{PowerEq5}) holds,
this does not mean that with high probability
\bee{PowerEq6}
T_{H_1}(n_1,n_2)  \le a(\de)
\ene

This is confirmed later based on a large number of Monte Carlo simulations.
Thus, in general the test in the form (\ref{PowerEq5}) does not guarantee
a high power of this test and this depends on the variance of $T_{H_1}(n_1,n_2)$.

\subsubsection{Simulation results}\label{Simulation1}

In Figure \ref{fig1}, the calculation of the number of centres
is based on the normal approximation using (\ref{NN}).
However, the required number of centres can be calculated
nearly exactly using Monte Carlo simulation of $10^6$ runs.

For this purpose we need to evaluate by simulation for every \\
$q = 0.5, 0.55, \dots, 0.85, 0.9$,
the minimal $N$
such that in (\ref{PowerEq6})
$$
\E [T_{H_1}(n_1,n_2) ]  \le \de
$$

Then the calculated nearly exact numbers of centres needed to detect the difference using criterion
(\ref{PowerEq5}) are (7, 8, 11, 14, 19, 28, 44, 78, 179), as reflected in Figure \ref{fig1}.
It's interesting to note that when we use these numbers,
the power of test defined in (\ref{PowerEq4})
varies in the range 0.7-0.73, so, not so large.

Correspondingly, to get a larger power, say, 80\%, the number of centres should be increased.
This will be explored further in the next part.

\subsubsection{Analysis of the power of the test $P_{Upp} \le \de$} 

Consider now the analysis of the power of the test $P_{Upp} \le \de$ using relation
(\ref{PowerEq4}).
Consider the same scenario as above,\\
$m_1=0.04$ (patients per centre per day); \\
$m_2 = qm_1$; $L_1=L_2=L$ and $L = 90$ (days);
$\de=0.1$.

Let us evaluate the power of the test $P_{Upp} \le \de$
for any given $q=0.5, 0.55,\dots, 0.85, 0.9$,
using a corresponding number of centres $N(q)$ from the vector (7, 8, \dots, 78, 179)
calculated as shown above using Monte Carlo simulation.

Consider Monte Carlo simulation using $10^6$ independent runs of three different vectors: \\
vector $vec.n_{1}$ of the number of patients $n_1$ in the 1st interval with mean rate $m_1$, \\
vector $vec.n_{2,H_0}$ of
the number of patients $n_2$ in the 2nd interval for hypothesis $H_0$ (simulation when $m_2=m_1$), \\
and vector $vec.n_{2,H_1}$ for hypothesis $H_1$, when $m_2=qm_1$.

As the result, we get three independently simulated vectors, $vec.n_{1}$, $vec.n_{2,H_0}$, $vec.n_{2,H_1}$.

First, we evaluate the value $a(\de)$ in the relation (\ref{PowerEq3})
using data set $ vec.n_{1},vec.n_{2,H_0}$ related to $H_0$.

Then, using the part of simulated data set related to $H_1$,
$vec.n_{1},vec.n_{2,H_1}$,
we evaluate the power of the test
\bee{PowerEq10}
\Pr (T_{H_1}(n_1,n_2) \le a(\de) )
\ene

Table \ref{Tab1} shows that for every $q$ the evaluated value $a(\de)$
in the column "Bound\_Type-I-error"
is very close to $\de = 0.1$, as expected.
The $P$-value for hypothesis $H_0$ is about $0.5$
and the $P$-value for hypothesis $H_1$ is about $0.1$, which is also expected.

Nevertheless, the power of this test is about 0.7 for all values $q$.
It is also expected that the power can be less than $1-\de = 0.9$.

\begin{table}
    \centering
    \caption{$P$-values and Power for different proportions of mean rates $m_1$, $m_2$}
    \begin{tabular}{ccccccccc}
        \toprule
        Proportion-q & Sites-Number & Bound\_Type-I-error & Pvalue\_H0 & Pvalue\_H1 & Power \\
        \midrule
        0.5  & 7  & 0.093 & 0.540   & 0.088  & 0.729 \\
        0.55 & 8  & 0.096 & 0.537  & 0.101  & 0.697 \\
        0.6  & 11 & 0.098 & 0.532  & 0.092  & 0.722 \\
        0.65 & 14 & 0.092 & 0.528  & 0.097  & 0.695 \\
        0.7  & 19 & 0.094 & 0.524  & 0.099  & 0.695 \\
        0.75 & 28 & 0.097 & 0.519  & 0.098  & 0.704 \\
        0.8  & 44 & 0.099  & 0.516  & 0.099  & 0.705 \\
        0.85 & 78 &  0.092 & 0.512 & 0.101 & 0.685 \\
        0.9 & 179 &  0.096 & 0.508 & 0.1   & 0.694 \\
        \bottomrule
    \end{tabular}
    \label{Tab1}
\end{table}

Therefore, this raises another question of interest.
What should be the minimal number of centres required
to reach a particular level of the power?
Let us consider a numeric solution to this problem for the case when required power is $80\%$.

Then, for any given $q=0.5, 0.55, \dots, 0.85, 0.9$,
according to (\ref{PowerEq4}), using Monte Carlo simulation with $10^6$ runs,
we need to find a minimal $N(q,2)$ such that the probability
\bee{PowerEq11}
\Pr (T_{H_1}(n_1,n_2) \le a(\de) ) \ge 80\%
\ene
where the values $(n_1,n_2)$ are simulated for a given scenario under $H_1$, so we use
vectors $vec.n_{1}$ and $vec.n_{2,H_1}$.

Table \ref{Tab2} shows the results, where the column "Sites-Number"
provides the required number of centres/sites to get power 80\%.

\begin{table}
    \centering
    \caption{The optimal number of centres/sites to reach power 0.8 for different proportions of mean rates $m_1$, $m_2$}
    \begin{tabular}{ccccccc}
        \toprule
        Proportion-q & Sites-Number & Bound\_Type-I-error & Pvalue\_H0 & Pvalue\_H1 & Power \\
        \midrule
        0.5  & 9   & 0.097 & 0.535 & 0.059 & 0.821 \\
        0.55 & 11  & 0.099 & 0.531 & 0.063 & 0.810 \\
        0.6  & 15  & 0.094 & 0.527 & 0.057 & 0.821 \\
        0.65 & 19  & 0.095 & 0.525 & 0.062 & 0.806 \\
        0.7  & 26  & 0.097 & 0.521 & 0.064 & 0.804 \\
        0.75 & 38  & 0.100 & 0.517 & 0.064 & 0.808 \\
        0.8  & 62  & 0.092 & 0.513 & 0.061 & 0.803 \\
        0.85 & 111 & 0.094 & 0.510 & 0.063 & 0.803 \\
        0.9  & 249 & 0.096 & 0.507 & 0.065 & 0.801 \\
        \bottomrule
    \end{tabular}
    \label{Tab2}
\end{table}

Here we can see again that the $P$-value for hypothesis $H_0$ is about $0.5$
as expected, however, the $P$-value for hypothesis $H_1$ is
now lower and is about $0.06$. This is also expected
as the number of centres is larger compared to the previous case,
thus, given hypothesis $H_1$, the $P$-value should be less.

\begin{figure}
    \centering
    \includegraphics[height=7cm, width=14cm]{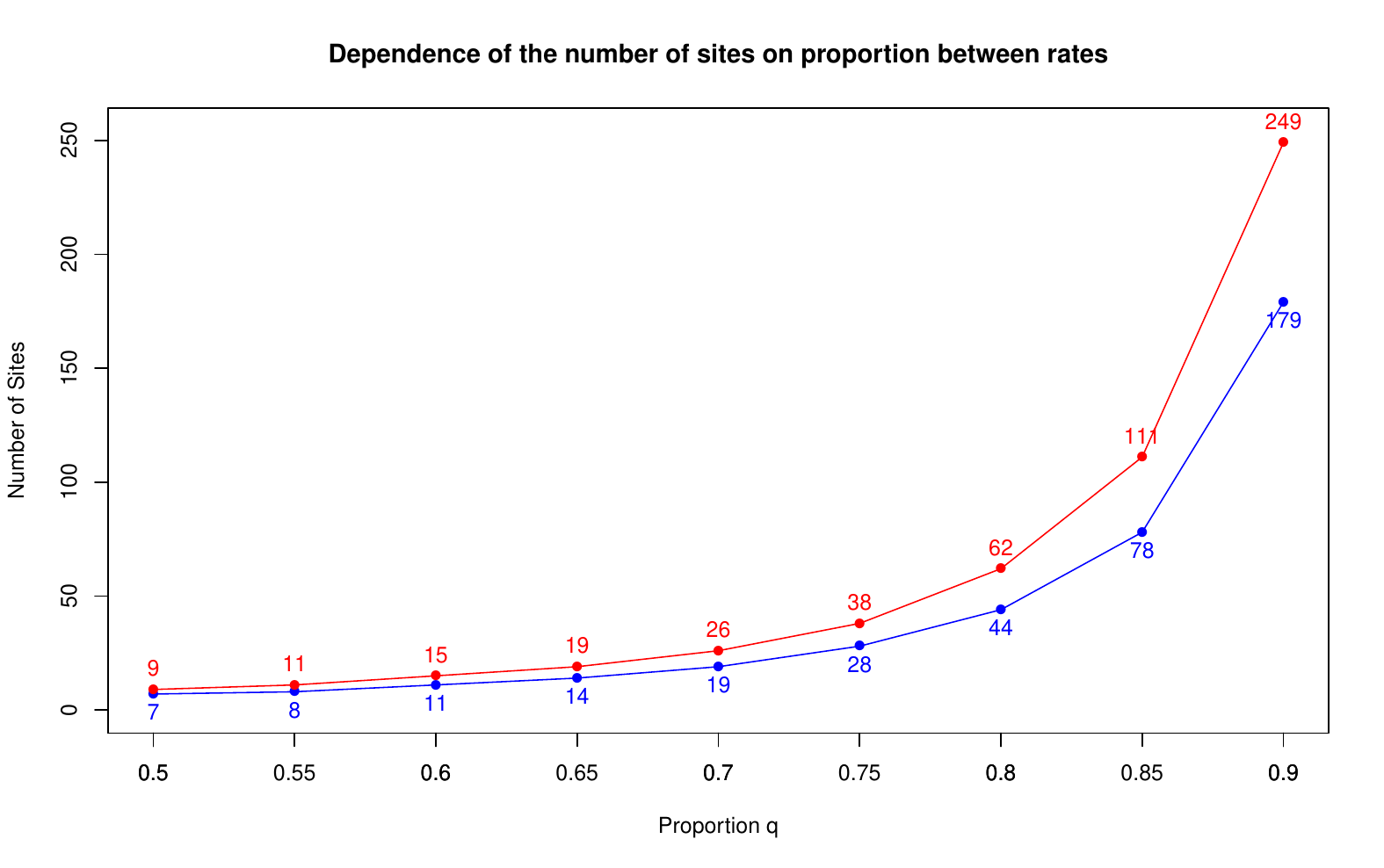}
    \caption{Comparing the number of centres/sites required to get $P_{Upp} \le 0.1$
    (blue line) with the number of centres/sites required to reach Power=0.8 (red line)
    for different proportion of mean rates $m_1$, $m_2$.}
       \label{fig2}
\end{figure}

Table \ref{Tab3} shows the comparison of the number of centres
for using $P$-value test and for using power of the test 80\%
using columns "Sites-Number" from tables \ref{Tab2} and \ref{Tab3}.
This is also shown visually in Figure \ref{fig2}.

Note that for a not so high proportion $q$, say $q \le 0.8$, these values don't differ much, though of course the numbers of centres required to reach a higher power are visibly larger.

\begin{table}
    \centering
    \caption{Comparing the number of centres/sites to get Upper Pvalue $\le 0.1$ and to reach Power=0.8 for different proportions of mean rates $m_1$, $m_2$}
    \begin{tabular}{cccccccccc}
        \toprule
      Proportion-q & 0.50 & 0.55 & 0.60 & 0.65 & 0.70 & 0.75 & 0.80 & 0.85 & 0.90 \\
        \midrule
        N-sites-Pvalue test & 7 & 8 & 11 & 14 & 19 & 28 & 44 & 78 & 179 \\
        N-sites-Power-0.8  & 9 & 11 & 15 & 19 & 26 & 38 & 62 & 111 & 249 \\
        \bottomrule
    \end{tabular}
\label{Tab3}
\end{table}

Nevertheless, when using test (\ref{PowerEq1}), in practice
if the test does not show a difference in rates, which means, $P$-value is not a small value,
we should not expect non-homogeneity in the rates.

However, if $P$-value is small, say $\le 0.1$,
according to power calculations, with a reasonable probability which can be even around 70\%,
we may expect that the mean rates in these two intervals are different.

In this case it can be recommended to consider a Poisson-gamma test
described in Section \ref{PG-test}
which is based on calculating $P$-values using a Poisson-gamma recruitment model.

\subsection{Poisson parametric test}

Let us consider another approach
which is based on using estimated
rates for a Poisson model in the total interval of observations
for calculating $P$-values for any given sub-interval.

Consider again two disjoint intervals $J_1 = [a,b]$ and $J_2 = [c,d]$ of length $L_1$ and $L_2$
with the number of active centres in these intervals $N_1$ and $N_2$.
Assume for simplicity that the centres are active during each interval.
Then the duration of the total recruitment window in interval $J_j$ is $U_j = N_jL_j, j=1,2$.

Assume Poisson model assumptions. Suppose the rates for each centre in these intervals are $m_1$ and $m_2$ (per centre per day), respectively.
That means, the number of recruited patients $n_j$ in interval $J_j$ is a Poisson random variable,
 $\pi_j = \Pi(m_j U_j)$.

Consider testing a hypothesis $H_0$ that $m_1=m_2=m$.

Denote the observed data (number of recruited patients) in each interval by $n_1$ and $n_2$
and put $n = n_1 + n_2$.

Then for a Poisson model, given $H_0$, the estimated rate $\what m$ in the union of the intervals
$J_1 \cup J_2$ is
\bee{Eq1}
\what m = \frac {n_1 + n_2}{U_1+U_2} =  \frac {n}{U}
\ene
where $U=U_1+U_2$.
Correspondingly, define for interval $J_1$ the lower and upper $P$-values as
\beeq{Eq2}
\begin{aligned}
& P_{Upp} = \Pr( \Pi( \what m U_1 ) \ge n_1 )  \\
& P_{Low} = \Pr( \Pi( \what m U_1 ) \le n_1 )
\end{aligned}
\eneq

These relations can be re-written in the form
\beeq{Eq3}
\begin{aligned}
& P_{Upp} = \Pr( \Pi( n p ) \ge n_1 ) \\
& P_{Low} = \Pr( \Pi( n p ) \le n_1 )
\end{aligned}
\eneq
where
$$
p = \frac {U_1}{U_1+U_2}
$$

It's interesting that for the previous non-parametric test, $P$-values have a similar form given in (\ref{PowerEq1}) where a binomial distribution is used instead of a Poisson one.
Note also that for a given $n$,
$$
\Va [\Bin(n,p)] = np(1-p) < \Va [\Pi(n p)] = n p
$$
Therefore, it is expected that the first non-parametric test is more powerful (requires lower
number of centres to detect a difference in means).

Note also that the approach in this section uses the estimated mean rate,
and this is another reason to expect that the first non-parametric test is more powerful.
This is supported by calculations below.

\subsubsection{Analysis of $P$-values}

Let us now investigate the dependence of $P$-values in (\ref{Eq3})
on $m_j$, $N_j$ and $L_j$ assuming the general case of possibly different
$m_1, m_2$. This will allow us to evaluate the number of centres needed for testing a particular difference in the rates $m_1, m_2$ and compare with the first test.

Note that
$$
P_{Upp} = \Pr( \Pi( n p ) \ge n_1 ) = \Pr( \Pi( n p ) - n_1\ge 0 )
$$

Consider now the expression
$X_2 = \Pi( (\pi_1+\pi_2) p ) - \pi_1$.
Then the calculations below follow similar steps as in Section \ref{Depend}.
As $\E[\pi_j ] = m_j U_j$,
\beeq{EX3}
\begin{aligned}
& \E[X_2] = \E[\Pi((\pi_1+\pi_2) p )] - \E[\pi_1] = \E[\pi_1+\pi_2]p - \E[\pi_1] \\
&= \E[\pi_1](p-1) + \E[\pi_2]p
 = \frac{U_1U_2}{U_1+U_2}(m_2-m_1)
\end{aligned}
\eneq

Note that
$\Va [\Pi( (\pi_1+\pi_2) p ) \mid \pi_1,\pi_2 ] = (\pi_1+\pi_2)p$,
$\Va [\pi_j ] = m_j U_j $. Therefore, the formula of total variance gives
\beeq{VarX3}
\begin{aligned}
& \Va[X_2] = \E[\Va[X_2 \mid \pi_1,\pi_2]] + \Va[\E[X_2 \mid \pi_1,\pi_2]] \\
& = \E[(\pi_1+\pi_2)p] + \Va[\pi_1(p-1)+\pi_2p] \\
& = (m_1U_1+m_2U_2)p + m_1U_1(p-1)^2 + m_2U_2p^2 \\
& = \frac{1}{U^2} \Big( m_1 U_1 (U_1 (U_1+U_2)+ U_2^2 ) + m_2 U_1U_2 (2 U_1+U_2 ) \Big)
\end{aligned}
\eneq

If to compare the expressions above for $\E[X_2]$ and $\Va[X_2]$ with
the expressions  (\ref{Mean}) and (\ref{TotVar})
for $\E[X]$ and $\Va[X]$ in the first Poisson
test, one can see that the mean is the same, however,
it's easy to calculate that
$$
\Va[X_2] - \Va[X] = \frac{1}{U^2}(U_1^3m_1 + U_1^2U_2m_2) > 0
$$

Denote
\bee{Eq8}
D = \sqrt{ m_1 U_1 (U_1 (U_1+U_2)+ U_2^2 ) + m_2 U_1U_2 (2 U_1+U_2 ) }
\ene

Now consider an approximation for $X_2$ by a normal distribution
similar to (\ref{NormApp}):
\beeq{Eq6}
\begin{aligned}
& X_2 \approx \E[X_2] + \sqrt{\Va[X_2]} \cdot Z \nn \\
& = \frac{U_1U_2}{U_1+U_2}(m_2-m_1) + \frac{D}{U_1+U_2} \cdot Z
\end{aligned}
\eneq
Then
\bee{PuppTwo}
P_{Upp} \approx \Pr( X_2 \ge 0) =
\Pr\left( Z \ge \frac{\frac{U_1U_2}{U_1+U_2}(m_1-m_2)}{\sqrt{\Va[X_2]}} \right)
\ene
and similarly for $P_{Low}$.

Note that for $m_1 > m_2$, as $\Va[X_2] > \Va[X]$,
the value $P_{Upp}$ in (\ref{PuppTwo}) above is greater than $P_{Upp}$ in the first Poisson
test, see (\ref{Pupp2}).
That means, using a parametric test we need to increase the values $U_1$ and $U_2$,
or correspondingly, increase the number of centres $N$, to reach the same level $\de$ compared to a non-parametric
test.


Consider now a particular case when $L_1=L_2=L$ and $N_1=N_2=N$.
Then using similar calculations we get the relation for $N$ required to reach
a particular confidence level $\de$:
\bee{NN8}
N = \frac{3 z_{\delta}^2}{m_1 L}\cdot\frac{1+q}{(1-q)^2}
\ene

Here it is clearly visible that  $N$ required for a parametric
test is greater compared to  (\ref{NN})
for a non-parametric one.

\subsection{Poisson-gamma test}\label{PG-test}

Consider an approach for testing the hypothesis $H_0$ that parameters $(\alpha,\beta)$ of the
$\text{PG}$ model are the
same for two disjoint intervals $[a,b]$ and $[c,d]$.
Assume there is possibly a different number of centres
in these intervals.

Consider combining data in both intervals.
Then, estimate parameters $(\alpha,\beta)$ of a PG model
using this data and maximum likelihood technique
described in \cite{anfed07,an11a}.
In each interval, using estimated parameters $(\hat{\alpha},\hat{\beta})$
and active recruitment windows, we can create the predictive $\text{PG}$ distribution.

Suppose we have some schedule of centre initiations for $N$ centres given by $(u_1,\dots,u_N)$.
Then we have $N_1$ centres active in interval $[a,b]$, and $N_2$ centres active in interval $[c,d]$ (not necessarily $N_1=N_2$). For each interval, say, $[a,b]$, define the duration
of the active recruitment window for centre $i$ using (\ref{e2}), as $v_i = u(a,b,u_i)$,
and correspondingly for interval $[c,d]$.

Denote the pools of active centres as $I_1$ and $I_2$ for intervals $j=1,2$, $[a,b]$ and $[c,d]$, respectively.
Then in interval $j$ we have the data $(k_i,v_i)$,
the number of patients recruited and the centre's active recruitment window.

Now, for every centre combine data from intervals $j=1,2$ to get data
$(k_i,v_i), i \in I_1 \cup I_2$,
the total number of patients and total active recruitment window which
is the sum of windows on both intervals (for simplicity keep the same notation
$k_i,v_i$ for the total patients and sum of windows).

We can use this data to estimate parameters of a $\text{PG}$ model, e.g.
using maximum likelihood technique and numerical optimisation
(assuming we have a suitable amount of data).
For data of $N$ centres, $\{(k_i,v_i),i=1,\dots,N\}$, the log-likelihood $\mathcal{L}(\alpha,\beta)$ has the form, \cite{anfed07,an11a}:
$$ 
\sum_{i=1}^N \left[\ln\Gamma(\alpha+k_i) + \alpha\ln\beta - \ln\Gamma(\alpha) - (\alpha+k_i)\ln(\beta+v_i) + k_i\ln v_i - \ln (k_i!) \right]
$$ 

Now, given estimated parameters $(\hat{\alpha},\hat{\beta})$
using the union of the intervals,
denote $m=\hat{\alpha}/\hat{\beta}$,
$s^2 = \hat{\alpha}/\hat{\beta}^2$.

In interval $j$ with pool of active centres $I_j$, consider the country mean and variance of the cumulative rate. So we have
$$
E(I_j) = m\sum_{i\in I_j} v_i, \quad S^2(I_j) = s^2 \sum_{i\in I_j} v_i^2
$$

Introduce the variables
$$A(I_j) = E^2(I_j)/S^2(I_j), \quad B(I_j) = E(I_j)/S^2(I_j)$$
In \cite{an-aus20} it was proved that the number of recruited patients $n_j$ in interval $j$
can be approximated by a PG variable
$\text{PG}(A(I_j),B(I_j))$ with parameters
$(A(I_j),B(I_j))$, where
$\text{PG}(A,B)$ is a mixed Poisson variable with the rate
${\rm Ga}(A,B)$.
Now we can define the upper and lower $P$-values in interval $j$ as
\beeq{pvalPG}
\begin{aligned}
P_{Upp}(I_j) = \Pr(\text{PG}(A(I_j),B(I_j)) \ge n_j )  \\
P_{Low}(I_j) = \Pr(\text{PG}(A(I_j),B(I_j)) \le n_j )
\end{aligned}
\eneq
where $n_j= \sum_{i\in I_j} k_i$ is the total number of patients recruited in the pool of centres $I_j$ in corresponding interval $j$.

To calculate these $P$-values,
we can use the cumulative probability function of a negative binomial distribution in R via the relation:
$$
\Pr(\text{PG}(A(I_j),B(I_j)) \le k ) =
\text{pnbinom}\left(k, \text{size}=\frac{E^2(I_j)}{S^2(I_j)},\text{prob}=\frac{E(I_j)}{E(I_j)+S^2(I_j)}\right)
$$

\subsubsection{Some analytic considerations}

To analyze the dependence on the number of centres
consider a particular case when
all centres are active during the whole intervals.
Define by $L_1$ and $L_2$ the durations of intervals. Then $ E(I_j) = m N_j L_j$, $ S^2(I_j) = s^2 N_j L_j^2 $,
so $ A(I_j) = \frac{m^2}{s^2}N_j, $ $ B(I_j) = \frac{m}{s^2 L_j} $,
and
\beeq{Eq10}
\begin{aligned}
P_{Upp}(I_j) = \Pr\left(\text{PG}\left(\frac{m^2}{s^2}N_j, \frac{m}{s^2 L_j} \right) \ge n_j \right) \\
P_{Low}(I_j) = \Pr\left(\text{PG}\left(\frac{m^2}{s^2}N_j,\frac{m}{s^2 L_j}\right) \le n_j \right)
\end{aligned}
\eneq
Now consider the behavior of values in (\ref{Eq10}) when hypothesis $H_0$ is not true.
That means we may have different parameters of a PG model $(m_j,s_j^2)$ in different intervals.
Note that the parameters $(m,s^2)$ in the relations above are estimated using data
in the union of intervals assuming $H_0$.
Now, as $\E [\Pi_\la(t)] = \E[\la] t$,
using the method of moments estimates, 
approximately,
\bee{Eq4}
m \approx \frac{n_1+n_2} {N_1 L_1+ N_2 L_2} = \frac{n_1}{N_1 L_1} \frac{N_1 L_1} {N_1 L_1+ N_2 L_2} +
\frac{n_2}{N_2 L_2} \frac{N_2 L_2} {N_1 L_1+ N_2 L_2}
\ene
Note that $n_j$ is a PG variable with parameters $(A_j,B_j)$,
where $A_j = \frac{m_j^2}{s_j^2} N_j, $ $  B_j = \frac{m_j}{s_j^2 L_j} $
and $ \E[n_j] = m_j N_j L_j $. Then,
$ m_j  \approx  \frac{n_j}{N_j L_j} $, and
from (\ref{Eq4}),
\beeq{Eq12}
m &\approx& m_1 \frac{N_1 L_1} {N_1 L_1+ N_2 L_2}  +
m_2 \frac{N_2 L_2} {N_1 L_1+ N_2 L_2}
\eneq
So, if $m_1 > m_2$, then the estimated value $m$ using a union of intervals satisfies
the relation $ m_1 > m > m_2 $.

However, under $H_0$, $\E [ \text{PG}(A(I_j),B(I_j)) ] = m N_j L_j$.
Thus, if $j=1$ in (\ref{Eq10}) for $P_{Upp}$,
using the means of the left and right-hand side parts
under the sign of probability we have
$$
\E [ \text{PG}(A(I_1),B(I_1)) ] = m N_1 L_1 < \E[n_1] = m_1 N_1 L_1.
$$
Thus, as $N_1 \to \infty$, $N_2 \to \infty$, the right-hand side
has a higher order of tendency to $\infty$, thus $P_{Upp} \to 0$,
and we can use this test to detect the case when $m_1 > m_2$.

\subsection{PG criterion power testing}

We follow the same methodology and scenario as given in Section \ref{PoisPower}.
Again, consider a scenario where: \\
$m_1=0.04$ (patients per centre per day); \\
$m_2 =q m_1$; $L_1=L_2=L$ and $L = 90$ (days);
$\de=0.1$. \\
Also, consider the shape parameter $\alpha=1/1.2^2$ for all centres in both intervals (recall that we are assuming $N_1=N_2=N$).

Now, using Monte Carlo simulation we need to find the minimal $N$ such that $\E [T_{H_1}(n_1,n_2) ] \le \de$. For a given value of $q$, this is calculated numerically by using a sequence of $N$ and selecting the smallest $N$ so that this inequality is satisfied. In calculation of $P$-values for the PG test, we use maximum likelihood approach for estimation of parameters over the joint intervals.

For each run of Monte Carlo simulation, interval data is generated using a Poisson-gamma process where $N$ gamma variables $vecla=(\lambda_1,\dots,\lambda_N)$
are simulated for $N$ different centres.
Then, in the first interval, we generate $N$ Poisson variables using recruitment rates $vecla$.
In the second interval, for hypothesis $H_0$, we generate data using the same rates.
For hypothesis $H_1$, in the second interval we use rates $q \times vecla$
to account for changing rate parameters.

\begin{figure}
    \centering
    \includegraphics[height=7cm, width=14cm]{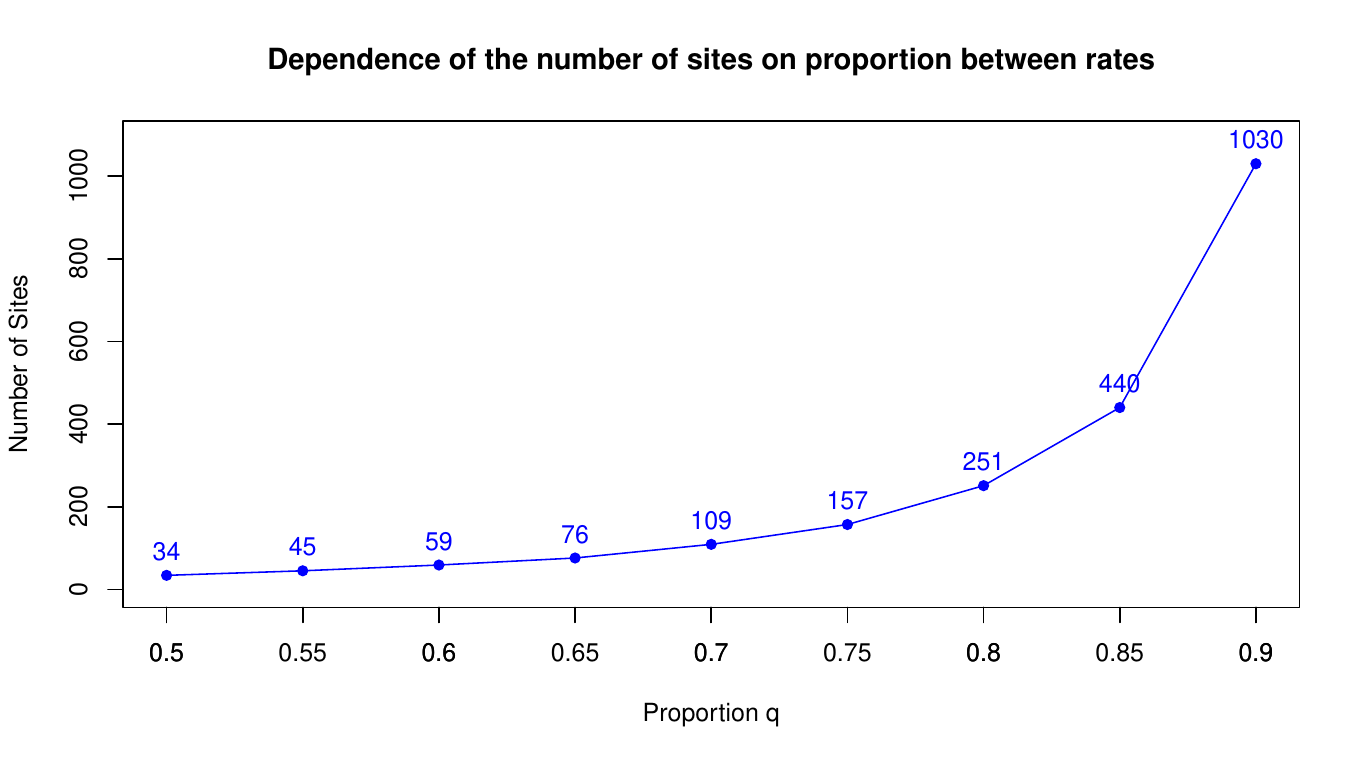}
    \caption{Dependence of the number of centres needed to detect the difference
    in rates by Monte Carlo simulation.}
    \label{fig3}
\end{figure}

For values $q=0.5,0.55,\dots,0.85,0.9$, Figure \ref{fig3} demonstrates
the dependence of $N$ on $q$ for this particular scenario.
Note that these values are close to exact values and will be even closer given a larger simulation number $>10^5$.

\subsubsection{Analysis of the power of the test $P_{Upp} \le \de$}

Consider the same scenario as above. Let us evaluate the power of the test $P_{Upp} \le \de$
for the vector of $q$ values $(0.5, 0.55, \dots, 0.85, 0.9)$,
using a corresponding number of centres $N(q)$ from the vector (34, 45,\dots, 440, 1030)
calculated as shown above by simulation.

Now we use Monte Carlo simulation with $10^5$ runs by simulating the number of patients in the first interval using rates $vecla$. In the second interval, use rates $vecla$ and $q \times vecla$, corresponding to hypotheses $H_0$ and $H_1$ respectively.
As the result we get three simulated vectors, $vec.n_{1},vec.n_{2,H_0},vec.n_{2,H_1}$, which are
the vectors of the number of patients recruited in each site.

First we evaluate the value $a(\de)$ in the relation $\Pr (T_{H_0}(n_1,n_2) \le a(\de) ) = \de$
using data set $vec.n_{1},vec.n_{2,H_0}$ related to $H_0$.

Then, using the part of simulated data set related to $H_1$,
$vec.n_{1},vec.n_{2,H_1}$,
we evaluate the power of the test $\Pr (T_{H_1}(n_1,n_2) \le a(\de) )$. Table \ref{Tab4} shows the results.

\begin{table}
    \centering
    \caption{$P$-values and Power for different proportions of mean rates $m_1$, $m_2$}
    \begin{tabular}{ccccccccc}
        \toprule
        Proportion-q & Sites-Number & Bound\_Type-I-error & Pvalue\_H0 & Pvalue\_H1 & Power \\
        \midrule
        0.5  & 34  & 0.066 & 0.517   & 0.098  & 0.657 \\
        0.55 & 45  & 0.061 & 0.517  & 0.098  & 0.647 \\
        0.6  & 59 & 0.063 & 0.514  & 0.099  & 0.655 \\
        0.65 & 76 & 0.066 & 0.513  & 0.101  & 0.653 \\
        0.7  & 109 & 0.066 & 0.512  & 0.099  & 0.663 \\
        0.75 & 157 & 0.068 & 0.509  & 0.100  & 0.665 \\
        0.8  & 251 & 0.070  & 0.507  & 0.100  & 0.669 \\
        0.85 & 440 &  0.072 & 0.504 & 0.102 & 0.664 \\
        0.9 & 1030 &  0.074 & 0.502 & 0.100   & 0.679 \\
        \bottomrule
    \end{tabular}
    \label{Tab4}
\end{table}

We can also provide a solution of this problem for the case of power equal to $80\%$. For this case, we must find minimal $N$ such that $\Pr (T_{H_1}(n_1,n_2) \le a(\de) ) \ge 1-\e$ with $\e=0.2$. Table \ref{Tab5} shows the results, where the column "Sites-Number" provides the required number of centres/sites to get $80\%$ power.

\begin{table}
    \centering
    \caption{The optimal number of centres/sites to reach power 0.8 for different proportions of mean rates $m_1$, $m_2$}
    \begin{tabular}{ccccccc}
        \toprule
        Proportion-q & Sites-Number & Bound\_Type-I-error & Pvalue\_H0 & Pvalue\_H1 & Power \\
        \midrule
        0.5  & 85   & 0.069 & 0.512 & 0.055 & 0.810 \\
        0.55 & 107  & 0.066 & 0.511 & 0.057 & 0.799 \\
        0.6  & 138  & 0.068 & 0.507 & 0.059 & 0.800 \\
        0.65 & 179  & 0.067 & 0.510 & 0.060 & 0.793 \\
        0.7  & 246  & 0.068 & 0.509 & 0.061 & 0.794 \\
        0.75 & 352  & 0.072 & 0.506 & 0.060 & 0.801 \\
        0.8  & 570  & 0.074 & 0.504 & 0.060 & 0.805 \\
        0.85 & 1002 & 0.073 & 0.504 & 0.061 & 0.799 \\
        0.9  & 2000 & 0.073 & 0.503 & 0.067 & 0.782 \\
        \bottomrule
    \end{tabular}
    \label{Tab5}
\end{table}

\begin{figure}
    \centering
    \includegraphics[height=7cm, width=14cm]{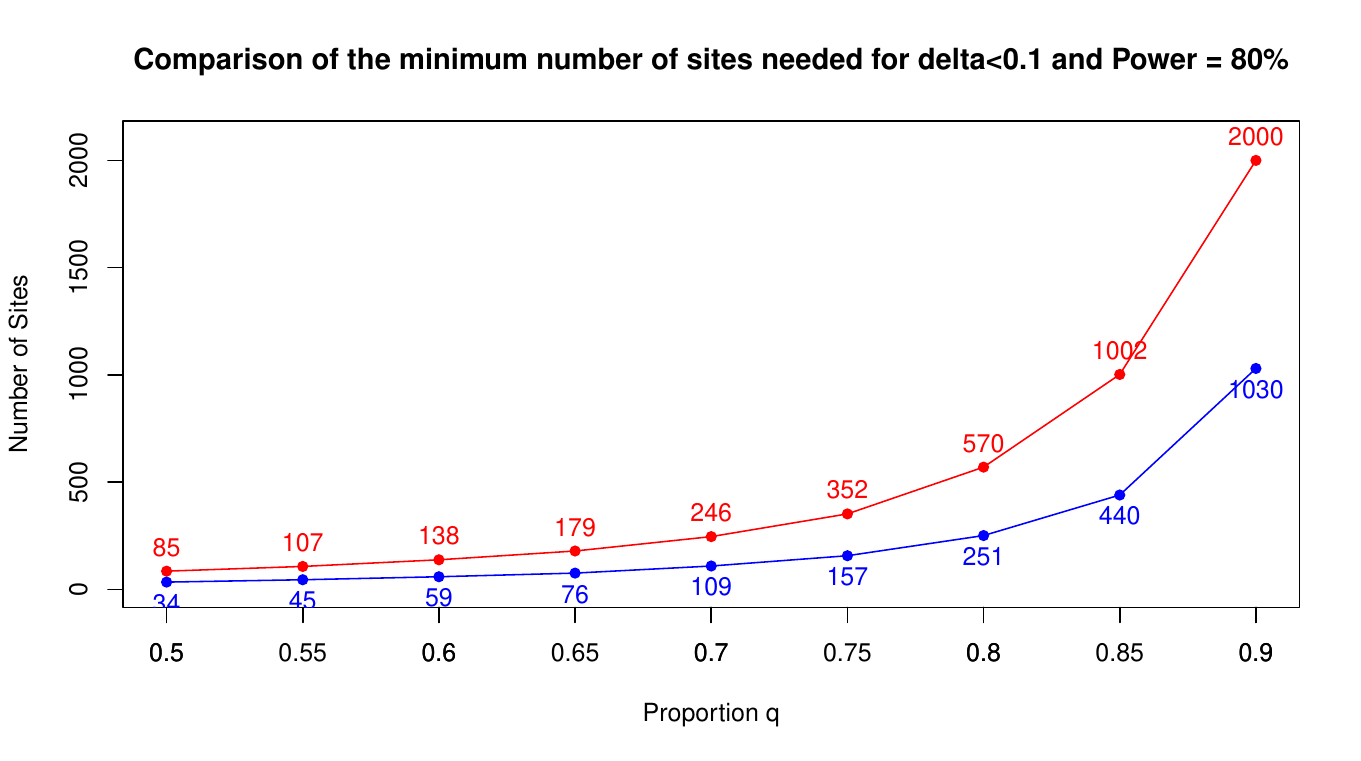}
    \caption{Comparing the number of centres/sites required to get Upper $P$-value $\le 0.1$
    (blue line) with the number of centres/sites required to reach Power of 0.8 (red line)
    for different proportion of mean rates $m_1$, $m_2$.}
       \label{fig4}
\end{figure}

Here we can see that the number of centres is much larger in order to obtain a power of $80\%$,
and the $P$-value for hypothesis $H_1$ is now much lower and is about $0.06$. Table \ref{Tab6}
shows the comparison of the number of centres for both approaches. Note that the difference in values
differs quite a bit, with more than $50$ extra centres needed for power of $80\%$.
This is also shown visually in Figure \ref{fig4}.

\begin{table}
    \centering
    \caption{Comparing the number of centres/sites to get upper $P$-value $\le 0.1$ and to reach
    power 0.8 for different proportions of mean rates $m_1$, $m_2$}
    \begin{tabular}{cccccccccc}
        \toprule
      Proportion-q & 0.50 & 0.55 & 0.60 & 0.65 & 0.70 & 0.75 & 0.80 & 0.85 & 0.90 \\
        \midrule
        N-sites-Pvalue test & 34 & 45 & 59 & 76 & 109 & 157 & 251 & 440 & 1030 \\
        N-sites-Power-0.8  & 85 & 107 & 138 & 179 & 246 & 352 & 570 & 1002 & 2000 \\
        \bottomrule
    \end{tabular}
\label{Tab6}
\end{table}

As this test involves estimating parameters, it
can be oriented to trials with quite large number of centres and patients recruited.
The properties of this test can be evaluated using Monte Carlo simulation.
Correspondingly, to detect the proportional difference in the mean rates,
it is required more centres compared to Poisson test.

Note that the Poisson test is more conservative as the random variation
in the number of recruited patients is larger for a PG model.
Therefore, it can be the case when Poisson test shows time-dependence,
but in reality this is just explained by random fluctuations.
In this case it is recommended to apply a PG test as the next step. However, if Poisson test doesn't show time-dependence, there is no need
to additionally use a PG test.

\section{Simulation example using non-homogeneous rates}\label{Sim-examples}

Consider simulating some artificial data in R using time-dependent rates as outlined in Section \ref{Simul}.
We compare a moving window approach with the classic approach using all data, where rate estimation
is performed using a maximum likelihood technique.
We also apply the Poisson and Poisson-gamma rate tests to the artificial data.

\begin{figure}
    \centering
    \includegraphics[height=7cm, width=14cm]{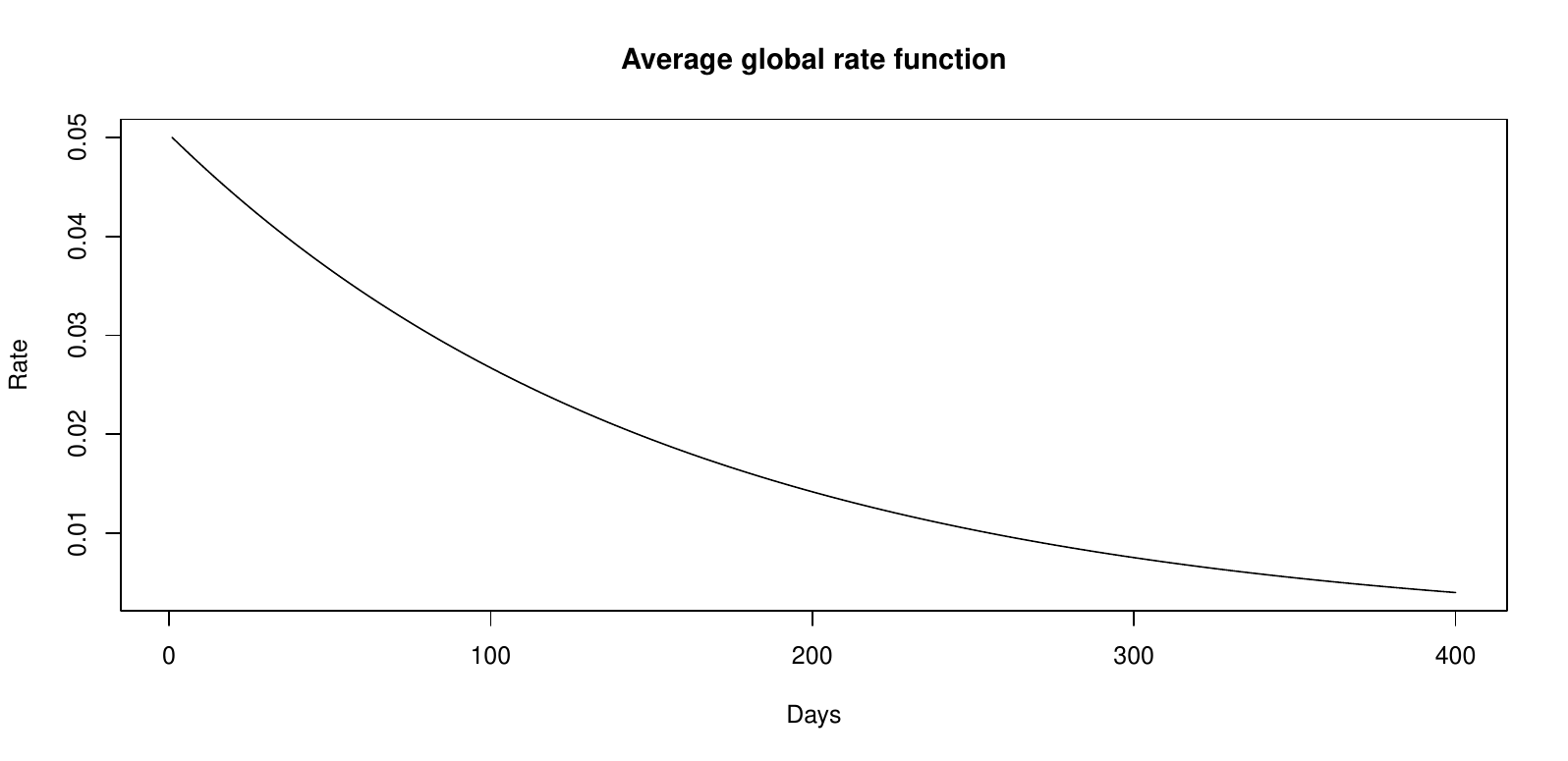}
    \caption{Average global rate function with exponential decay.}
       \label{fig5}
\end{figure}

\newpage
Consider the following parameters of simulation:

\begin{itemize}
\item[--] 200 centres where each centre has mean rate 0.02 and coefficient of variation of 1.2. So, for each centre $i$, $\alpha_i = 1/1.2^2$ and $\beta_i = \alpha_i/0.02$

\item[--] Centres  activated within 4 months using uniform grid

\item[--] In each centre $i$, the rate $\lambda_i r(t)$ has the following form: \\
Exponential decrease over 400 days from $2.5\times$ (the mean rate) to $0.2\times$ (the mean rate) (shown in Figure \ref{fig5})

\item[--] Recruitment target of 1000 patients to be recruited in 400 days
\end{itemize}

Then, a time-dependent PG process can be simulated using technique described in Section \ref{Simulation}
and we can perform rate testing at some interim time, specifying some intervals and interval width to test on.

Consider simulating data for 400 days, ensuring recruitment target is reached, and then consider an interim time of 200 days.

We can look at the average global rate over time. If each centre $i$ has simulation rate parameters $(\alpha_i$, $\beta_i)$
and rate function $r_i(t)$, then the mean rate at time $t$ is $\frac{\alpha_i}{\beta_i}r_i(t)$.
Then we can average over all centres at time $t$, as shown in Figure \ref{fig5}.
In the following simulation, $r_i(t)$ and parameters $(\alpha_i, \beta_i)$ are the same across all centres.

\begin{figure}
    \centering
    \includegraphics[height=7cm, width=14cm]{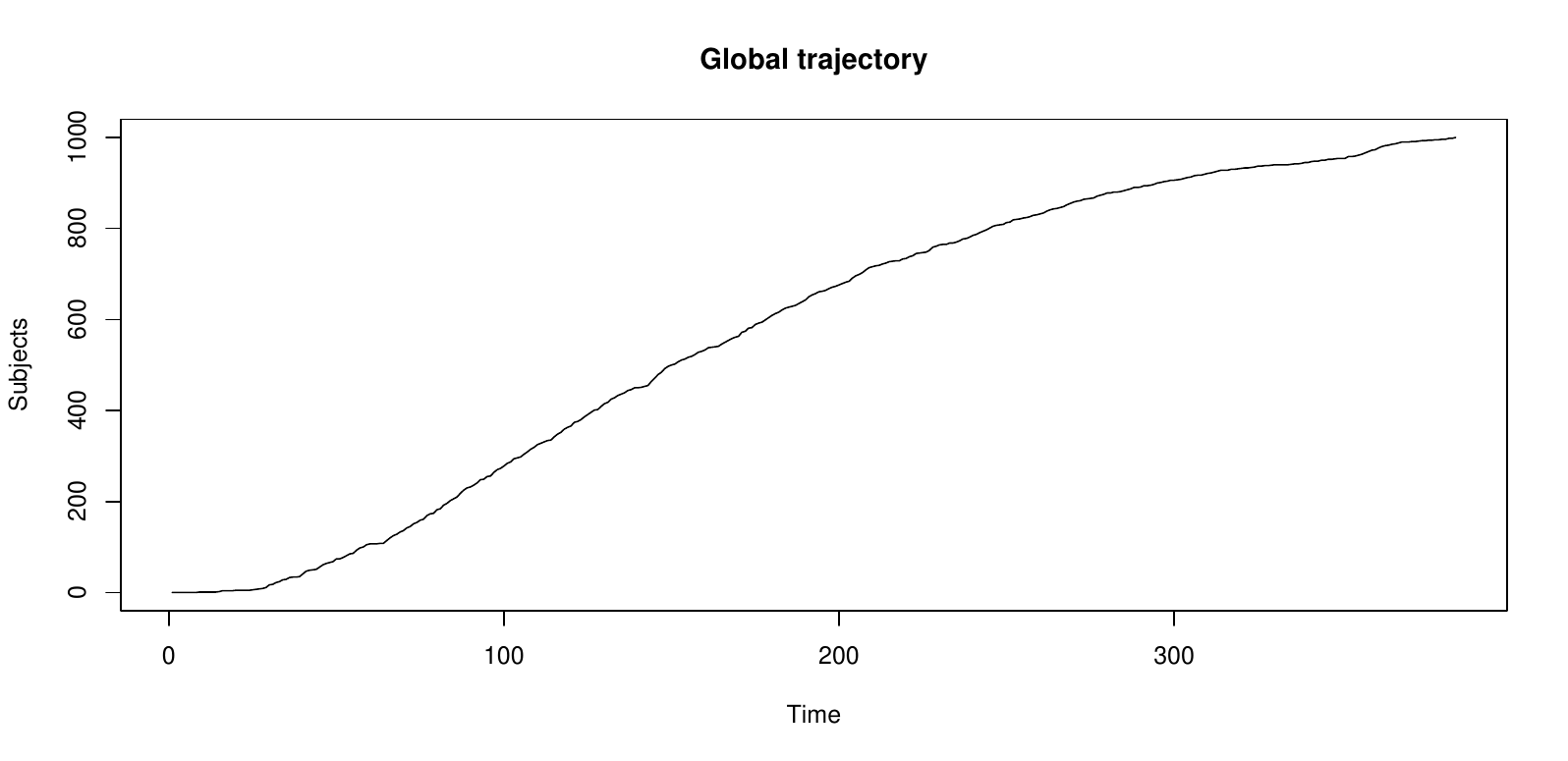}
    \caption{Simulated trajectory of recruitment in all centres with time in days.}
       \label{fig6}
\end{figure}

The simulated trajectory of recruitment is shown in Figure \ref{fig6}. Note that the global target of 1000 was reached on day 384.

Given an interim time of 200 days, we can perform rate testing on two adjacent, disjoint intervals of 60 days.
Both Poisson and Poisson-gamma tests were able to detect a declining rate between
days 80-140 and 140-200 using the simulated data.

Now, given a detected rate change, we can use a moving window approach, where we use only the latest 60 days of data
and maximum likelihood technique for estimating rate parameters for future prediction.

Consider two approaches for prediction: using all data from the start and using only the latest 60 days of data.

The results using all data were an estimated mean completion date on day 269 with lower and upper bounds of 263 and 274. This gave a probability of success of 1.
The results using only the most recent 60 days of data gave an estimated mean completion date on day 287, with lower and upper bounds of 279 and 295, and also a probability of success of 1.
Lower and upper bounds are calculated with 80\% confidence interval.

\begin{figure}
    \centering
    \includegraphics[height=7cm, width=14cm]{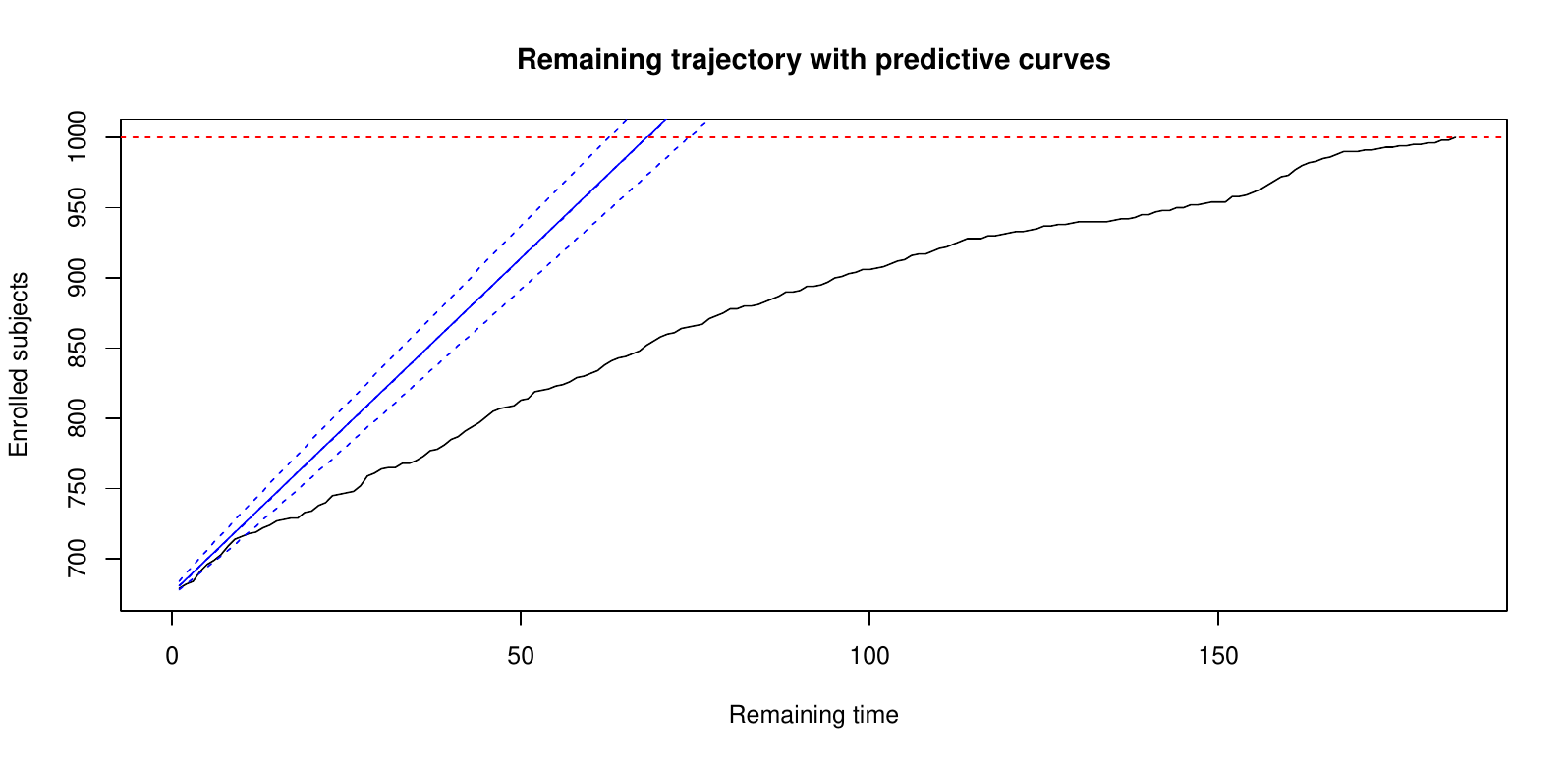}
    \caption{Reprojection at 200 days using all data and maximum likelihood technique. Reprojected mean and bounds in blue, global target in red.}
       \label{fig7}
\end{figure}

\begin{figure}
    \centering
    \includegraphics[height=7cm, width=14cm]{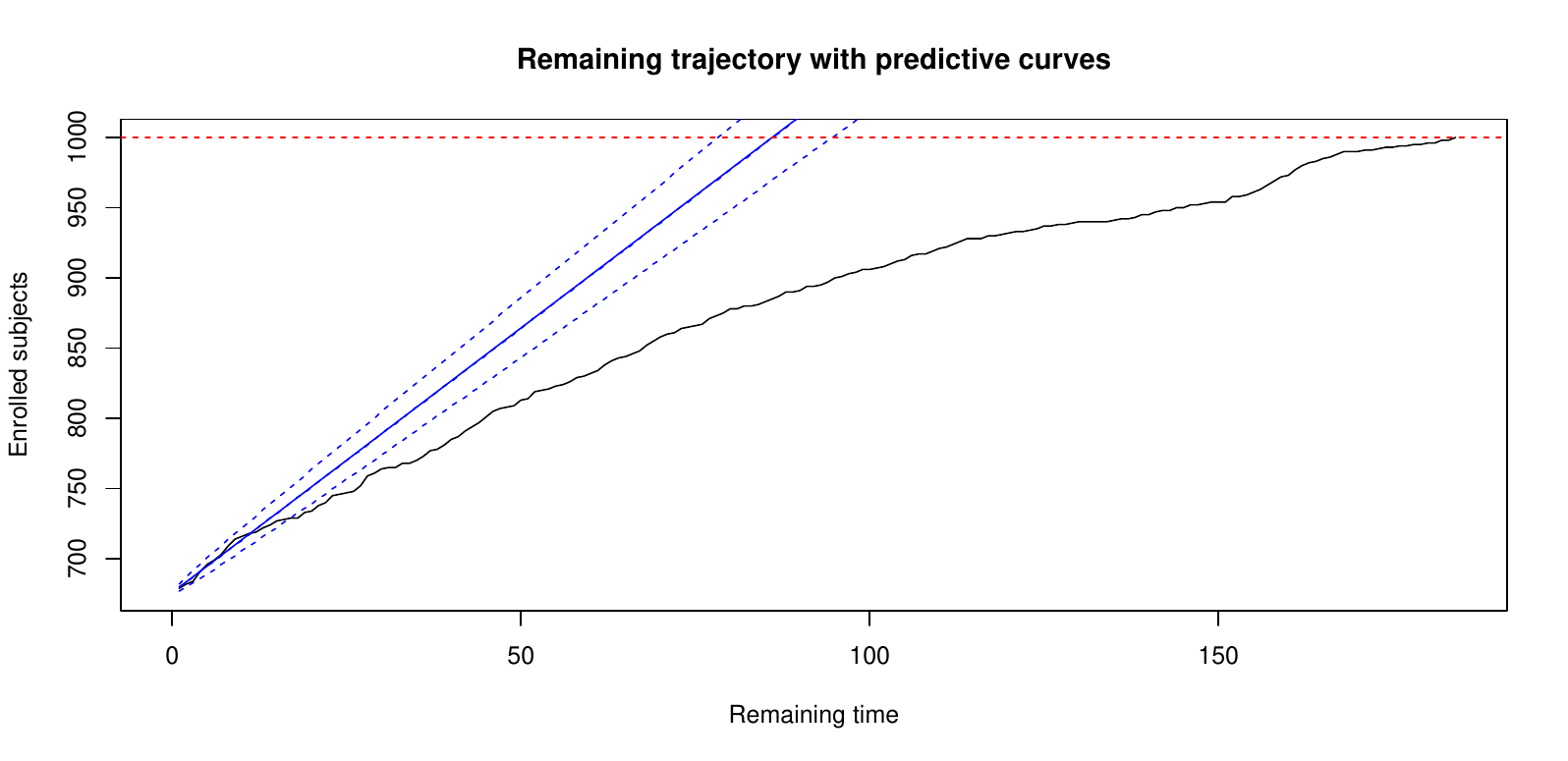}
    \caption{Reprojection at 200 days using most recent 60 days of data and standard maximum likelihood technique. Reprojected mean and bounds in blue, global target in red.}
       \label{fig8}
\end{figure}

The results are shown visually in Figure \ref{fig7} using all data, and Figure \ref{fig8} using the moving window approach.

It can be seen that using a moving window approach after rate-testing detects a change can give slightly more accurate prediction and is better suited to recent data. This approach is an improvement upon the standard approach when rates are declining steadily. However, if rates continue to decline then predictions can be poor as future time-dependence and rate decline is not considered.

Note that using a smaller window of data may improve prediction, but rate tests may not be able to successfully detect a changing rate using small intervals for testing. Also, rate estimation in a small window may give unreliable estimates when the amount of data is small.

A more advanced approach can be to use the methodology outlined in Section \ref{Non-Hom}. First, assuming the rate function $r(t)$ is known, we can apply a maximum likelihood approach as in Section \ref{Time-dep-estim}. For our simulated data, this gave a estimated mean completion data on day 391 with lower and upper bounds of 363 and 425. Here, the probability of success to complete recruitment in 400 days was 0.645.

\begin{figure}
    \centering
    \includegraphics[height=7cm, width=14cm]{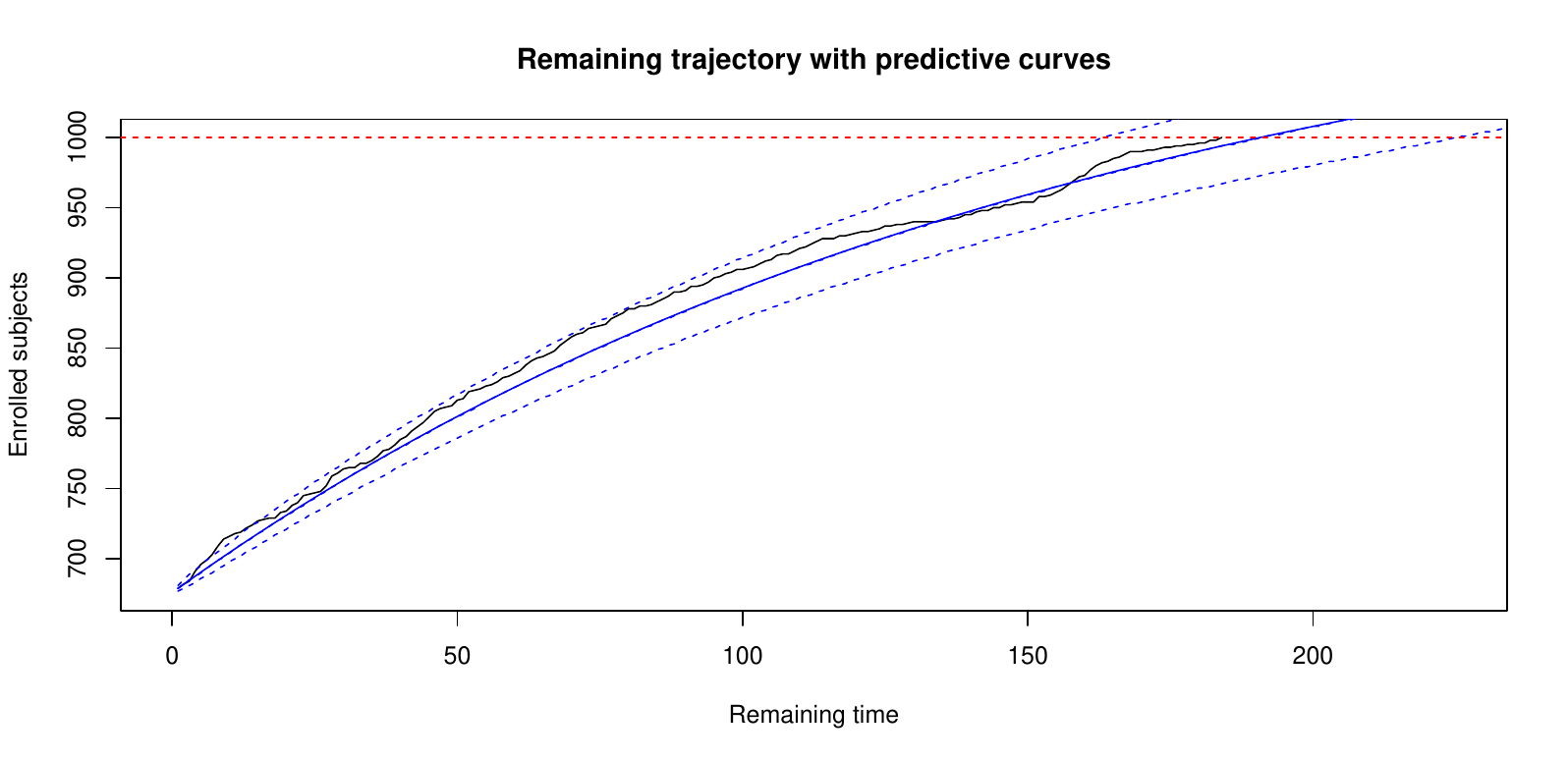}
    \caption{Reprojection at 200 days using time-dependent maximum likelihood technique, assuming $r(t)$ is known. Reprojected mean and bounds in blue, global target in red.}
       \label{fig9}
\end{figure}

The results are shown visually in Figure \ref{fig9}. It is clear that the approach
when $r(t)$ is known yields superior predictions over the previous approaches.
However, in reality the function $r(t)$ is not known,
so the main task is to suitably approximate the rate function and estimate the form of the
function through the use of previous data.

Similar scenarios can also be considered, where recruitment is coming from centres in countries with different rate parameters and forms of rate function $r(t)$. The conclusions will be similar to the above.

\section{Implementations}\label{Implem}

To verify how the technique for modelling and forecasting non-homogeneous patient recruitment in clinical trials is working, there were considered different case studies
using artificial data and exhaustive Monte Carlo simulation.

At the initial stage when we only have planned rates and some schedule
of centre initiation, a few specific forms of function
$r(t)$ including piece-wise linear dependence and exponential decay over time were considered.
Recruitment was simulated using technique proposed in Section \ref{Simulation}
and prediction of the recruitment was created using the analytic technique
proposed in Section \ref{Non-Hom}.
The results shows a perfect coincidence of the analytically calculated
mean and predictive bounds with the results of simulation
even for not so large number of clinical centres.

At the interim stage, first it is recommended to test the recruitment rates
on time-dependence using two previous time-intervals
and the developed Poisson and PG criteria.
If the tests do not show any time-dependence, for predicting
future recruitment we can use a standard  PG model and the techniques developed in \cite{an11a}.

However, if the tests show time-dependence, there can be different options.
One way can be to use a window of the most recent data,
e.g. 3 months, and create predictions using the same technique as in
\cite{an11a}.
In this case, the recruitment rates will use only the latest
3-months of data and will be adjusted to the time-trend
by repeating this procedure at every interim time in the future. Results from simulation show that
when rates are changing monotonically, this approach is an improvement over the techniques in \cite{an11a} .

Another more advanced opportunity can be to assume some particular time-dependence,
say, linear or exponential function $r(t)$. Then, we can estimate parameters
of the rates and of the function $r(t)$ using data in moving window and maximum likelihood technique, and then
use these parameters for predicting future recruitment using
the analytic technique proposed in Section \ref{Non-Hom}.

The results of simulation of several scenarios show
that if the rate is steadily changing over time
(linear or exponential increase/decrease), both approaches show improved
results compared to using a standard PG model,
where the second approach is also outperforming the first one.

\section*{Acknowledgment}

The authors would like to express their sincere gratitude to
Matt Austin, David Edwards, Behzad Beheshti
and the whole Data Science Team at Center for Design \& Analysis, Amgen for
constant and useful discussions on forecasting
patient recruitment in multicentre clinical trials.


\begin{thebibliography}{99}

\bibitem{anfed07}
V. Anisimov  and V. Fedorov, Modelling, prediction and adaptive
adjustment of recruitment in multicentre trials, {\em Statistics in Medicine},
26, 27, 2007, pp. 4958--4975.

\bibitem{anfed07_Nant}
V. Anisimov  and V. Fedorov, Design of multicentre clinical trials with random enrolment,
Chapter 25, in book {\em Advances in Statistical Methods for the Health Sciences}; Series:
Statistics for Industry and Technology,  Balakrishnan, N.; Auget, J.-L.; Mesbah, M.; Molenberghs, G.,
Birkh${\rm \ddot{a}}$user Boston, 2007, pp. 387--400.

\bibitem{an-dow-fed07}
V. Anisimov, D. Downing  and  V. Fedorov,  Recruitment in multicentre trials:
prediction and adjustment, in
{\em mODa 8 - Advances in Model-Oriented Design and Analysis}, 2007, pp. 1--8.

\bibitem{an09a}
V. Anisimov, Predictive modelling of recruitment and drug
supply in multicenter clinical trials. In: {\em Proc. of the Joint
Statistical Meeting}, Biopharmaceutical Section, Washington, DC,
American Statistical Association, 2009, pp. 1248--1259.

\bibitem{an11a}
V. Anisimov,  Statistical modeling of clinical trials (recruitment and randomization),
{\em Communications in Statistics - Theory and Methods}, 40, 19-20, 2011, pp. 3684--3699.

\bibitem{an11b}
V. Anisimov, Predictive event modelling in multicentre clinical trials with waiting time to response,
{\em Pharmaceutical Statistics}, v. 10, iss. 6, 2011, pp. 517-522.

\bibitem{an16b}
V. Anisimov,  Discussion on the paper "Real-time prediction of clinical trial enrollment
and event counts: a review" by D.F. Heitjan et al., {\em Contemporary Clinical Trials},
40, 2016, pp. 7--10.


\bibitem{an20}
V. Anisimov,  Modern analytic techniques for predictive modelling of
clinical trial operations, Chapter 8, in book {\em Quantitative Methods in
Pharmaceutical Research and Development: Concepts and Applications},
O. Marchenko, N. Katenka (Eds),
Springer International Publ., 2020, pp. 361--408.


\bibitem{an-aus20}
V. Anisimov and  M. Austin,
Centralized statistical monitoring of clinical trial enrollment performance,
{\em Communications in Statistics - Case Studies and Data Analysis}, 6, 4, 2020,
pp. 392--410.

\bibitem{an-aus22a}
V. Anisimov and M. Austin,
Modeling restricted enrollment and optimal cost-efficient design in multicenter clinical trials,
{\em arXiv:2212.12930}, 2022, 22 pages.

\bibitem{an-StRC22}
V. Anisimov, S. Gormley, R. Baverstock and C. Kineza,
Predicting event counts in event-driven clinical trials accounting for cure and ongoing recruitment,
Chapter 9, in book "Data Analysis and Related Applications 2. Multivariate, Health and Demographic Data Analysis",
V. 10 - Big Data, Artificial Intelligence and Data Analysis,
Edited by K. N. Zafeiris, C. H. Skiadas, Y. Dimotikalis, A. Karagrigoriou, C. Karagrigoriou-Vonta,
Wiley \& ISTE, Aug 2022, pp. 121-142.

\bibitem{an-aus23}
V. Anisimov and M. Austin,
Forecasting and optimizing patient enrolment in clinical trials under various restrictions,
chapter 23,
in book: {\em Stochastic Processes, Statistical Methods, and Engineering Mathematics},
Springer Proceedings in Mathematics \& Statistics 408, 2023, pp. 511--540.

\bibitem{an-23}
V. Anisimov,
An analytic methodology for forecasting patient enrolment performance in multicentre clinical trials, ICoMS '23: Proceedings of the 2023 6th International Conference on Mathematics and Statistics, July 14-16, 2023, Leipzig, Germany. ACM, New York, NY, USA, 2023, pp. 91-97.

\bibitem{an-24}
V. Anisimov,
Modelling/forecasting patient recruitment in clinical trials using Poisson-gamma model with time-dependent rates, ICoMS '24: Proceedings of the 2024 7th International Conference on Mathematics and Statistics", June 23-25, 2024, Amarante, Portugal, ACM, New York, NY, USA, 2024, (to appear).

\bibitem{baksenn13}
 A. Bakhshi, S. Senn and A. Phillips, Some issues in predicting patient
  recruitment in multi-centre clinical trials. {\em Statistics in
  Medicine}, 32(30), 2013, pp. 5458--5468.


\bibitem{barnard10}
K.D. Barnard, L. Dent  and  A. Cook,  A systematic review of models to predict
recruitment to multicentre clinical trials,
{\em BMC Medical Research Methodology}, 10, 63, 2010.

\bibitem{bernardo04}
J.M. Bernardo  and  A.F.M. Smith, {\em Bayesian Theory}, John Wiley \&  Sons:
Hoboken, NJ, USA, 2004.

\bibitem{Perevoz2022}
N. Best, I. Perevozskaya, D. Lunn, G. Archer, J. Euesden, C. Fillmore, V. Sherina,
D. Thompson and M. Zwierzyna, 'Predicting the COVID-19 pandemic impact on clinical trial recruitment',
Statistics in Biopharmaceutical Research,
14(1), 2022, pp. 67--79.

\bibitem{carter05}
R.E. Carter, S.C. Sonne and K.T. Brady,
Practical considerations for  estimating clinical trial accrual periods: Application to a
  multi-center effectiveness study, {\em BMC Medical Research
  Methodology}, 2005, 5, pp. 11--15.

\bibitem{gajew-sim08}
B.J. Gajewski, S.D. Simon and S.E. Carlson, 
  Predicting accrual in   clinical trials with Bayesian posterior predictive distributions.
  {\em Statistics in Medicine},
  27, 2008, pp. 2328--2340.


\bibitem{gkioni19}
E. Gkioni, R. Riusd, S. Dodda and C. Gamblea,
A systematic review describes models for recruitment prediction at the
design stage of a clinical trial, {\em Journal of Clinical
Epidemiology}, 115, 2019, pp. 141--149.

\bibitem{heitjan15}
D.F. Heitjan, Z. Ge  and  G.S. Ying,  Real-time prediction of clinical trial
enrollment and event counts: a review, {\em Contemporary Clinical  Trials},
45, part A, 2015, pp. 26--33.

\bibitem{lan-heitjan19}
Y. Lan, G. Tang and D.F. Heitjan,
Statistical modeling and prediction of clinical trial recruitment.
 {\em Statistics in Medicine}, 38(6), 2019, pp. 945-955.

\bibitem{savy12}
  G. Mijoule, S. Savy and N. Savy,
  Models for patients' recruitment in  clinical trials and sensitivity analysis,
  {\em Statistics in Medicine},
  31(16), 2012, pp. 1655--1674.

\bibitem{savy17}
M.N. Minois, V. Lauwers-Cances, S. Savy, M. Attal, S. Andrieua,
V. Anisimov and  N. Savy, 
Using Poisson-gamma model to evaluate the duration of
  recruitment process when historical trials are available.
  {\em Statistics in Medicine},
  36(23), 2017, pp. 3605--3620.

\bibitem{Perper2023}
A. Perperoglou, Y. Zhang, and D-K. Kipourou,
Modeling time-varying recruitment rates in multicenter
clinical trials, Biometrical J., v. 65, iss. 6, Aug 2023, pp. 1--12.

\bibitem{senn97}
S. Senn, 
{\em Statistical Issues in Drug Development}. Wiley:  Chichester, 1997.

\bibitem{senn98}
S. Senn,  Some controversies in planning and analysis  multi-center trials.
{\em Statistics in Medicine}, 17, 1998, pp. 1753--1756.


\bibitem{Savy2023}
A. Turchetta, N. Savy, D. A. Stephens, E. E.M. Moodie, and M. B. Klein,
A time-dependent Poisson-Gamma model for recruitment forecasting in multicenter studies,
Statistics in Medicine, 42, 2023, pp. 4193--4206.

\bibitem{Armando2024}
A. Turchetta, E. E.M. Moodie, D. A. Stephens, N. Savy, and Z. Moodie, 
The time-dependent Poisson-gamma model in practice: Recruitment forecasting in HIV trials,
Contemporary Clinical Trials, June, 2024.

\bibitem{urban2022}
S. Urbas, C. Sherlock, and P. Metcalfe,
Interim recruitment prediction for multi-center clinical trials,
Biostatistics, 2022, pp. 85--506.

\end{thebibliography}
\end{document}